\pdfoutput=1

\documentclass[tinyml]{acmart}

\usepackage{amsmath}
\usepackage{amsthm}
\usepackage{algorithm}
\usepackage{algorithmic}
\usepackage{xspace}
\usepackage{graphicx}
\usepackage{subfigure}
\usepackage{wrapfig}
\usepackage{graphicx}
\usepackage{subfigure}
\usepackage{url}
\usepackage{color,xcolor}
\usepackage{multirow}
\usepackage{gensymb}
\usepackage{longtable}
\usepackage{arydshln}
\usepackage{tikz}
\usepackage{comment}
\usepackage{dsfont}
\usepackage{framed}
\usepackage{enumitem}
\usepackage{xspace}
\usepackage{bm}
\usepackage{multirow}
\usepackage{adjustbox}

\usepackage{comment}
\usepackage[algo2e]{algorithm2e}
\usepackage{soul, xcolor}

\newcommand{\note}[1]{{\color{black}{\textbf{}#1\textbf{}}}}

\setstcolor{red}

\newcommand{\ouralg}{semi-decoupled\xspace}
\newcommand{\design}{hardware-software co-design\xspace}

\newcommand{\tabincell}[2]{\begin{tabular}{@{}#1@{}}#2\end{tabular}}

\AtBeginDocument{%
	\providecommand\BibTeX{{%
			\normalfont B\kern-0.5em{\scshape i\kern-0.25em b}\kern-0.8em\TeX}}}

\AtBeginDocument{%
  \providecommand\BibTeX{{%
    \normalfont B\kern-0.5em{\scshape i\kern-0.25em b}\kern-0.8em\TeX}}}

\setcopyright{rightsretained}
\copyrightyear{2022}
\acmYear{2022}

\begin{document}

\title{A Semi-Decoupled Approach to Fast and Optimal Hardware-Software Co-Design of Neural Accelerators}

\author{Bingqian Lu}
\affiliation{%
	\institution{UC Riverside}
}

\author{Zheyu Yan}
\affiliation{%
	\institution{Notre Dame}
}

\author{Yiyu Shi}
\affiliation{%
	\institution{Notre Dame}
}

\author{Shaolei Ren}
\affiliation{%
	\institution{UC Riverside}
}

\renewcommand{\shortauthors}{Bingqian Lu, et al.}

\begin{abstract}
	In view of the performance limitations
	of fully-decoupled designs for neural architectures
	and accelerators, hardware-software co-design
	has been emerging to fully
	reap the benefits of flexible design spaces and optimize neural
	network performance.
	Nonetheless,
	such co-design also  enlarges the total search space to practically infinity
	and presents substantial challenges.
	While the prior studies have been focusing
	on improving the search efficiency
	(e.g., via reinforcement learning),
	they commonly rely on co-searches over the entire
	architecture-accelerator design space.
	In this paper, we propose a \emph{semi}-decoupled approach
	to reduce
	the size of the total design space by orders of magnitude, yet without
	losing optimality.
	We first perform neural architecture search
	to obtain a small set of optimal
	architectures for one accelerator candidate. Importantly, this
	is also the set of (close-to-)optimal architectures for
	other accelerator designs based on the property
	that neural architectures' ranking orders in terms of inference latency and energy consumption on different accelerator designs are highly similar.
	Then, instead of considering all the possible
	architectures, we optimize the accelerator design only
	in combination
	with this small set of architectures, thus significantly reducing
	the total search cost.
	We validate our approach by conducting experiments on various
	architecture spaces for accelerator designs with different dataflows.
	Our results highlight that we can  obtain
	the optimal design by only navigating over the reduced search space.
	The source code of this work is at \note{\url{https://github.com/Ren-Research/CoDesign}}.
\end{abstract}


\maketitle

\pdfoutput=1

\section{Introduction}
\label{sec: introduction}

Neural architecture search (NAS) has been commonly used as a powerful tool to automate the design of  efficient deep neural network (DNN) models \cite{zoph_nas_with_reinforcement_learning}.
As DNNs are being deployed on increasingly diverse devices such
as tiny Internet-of-Things devices,
state-of-the-art (SOTA) NAS is turning
hardware-aware by
further taking into consideration the target hardware as a crucial
factor that affects the resulting performance (e.g., inference latency) of NAS-designed models \cite{DNN_NAS_HardwareNAS201_Benchmark_Rice_ICLR_2021_li2021hwnasbench,DNN_AccuracyPrediction_NeuralPredicto_YiranChen_Duke_2019_wen2019neural,weight_sharing_perform_random_search_CVPR, multi_hardware_mobile_models_NAS_CVPR, DNN_NAS_MnasNet_Google_CVPR_2019_Tan_2019,DNN_NAS_EfficientNet_Google_ICML_2019_pmlr-v97-tan19a,DNN_FBNet_HardwareAwareConvNetDesign_CVPR_2019_Wu2018FBNetHE}

Likewise, optimizing hardware accelerators
built on Field Programmable Gate Array (FPGA) or Application-Specific Integrated Circuit (ASIC), as well as
the corresponding dataflows (e.g.,
 scheduling DNN computations
 and mapping them on hardware),
is also critical for speeding up DNN execution 
\cite{DNN_NAS_MultiFPGA_YiyuShi_Journal_TECS_2019_Jiang:2019:ASS:3365919.3358192,DNN_AutoDNNChip_FPGA_ASIC_YingyanLin_Rice_DemingChen_UIUC_FPGA_2020_10.1145/3373087.3375306,DNN_NAS_CompilierDesign_UCSD_ICLR_2020_Ahn2020Chameleon:}.

While both NAS and accelerator optimization can effectively improve
the DNN performance (in terms of, e.g., accuracy and latency),
they are traditionally performed in a siloed manner, without
fully unleashing the potential of design flexibilities.
As shown in recent
studies \cite{DNN_NAS_Hardware_YiyuShi_ICCAD_2019_Jiang:2019:AVE:3316781.3317757,song_han_neural_accelerator_architecture_search}, such a decoupled approach does not explore potentially better
combinations of architecture-accelerator designs, leading to highly sub-optimal DNN performance.
As a result,
co-design of neural architectures and accelerators (a.k.a.,
\design) has been emerging to discover jointly optimal architecture-accelerator designs
\cite{DNN_AutoDNNChip_FPGA_ASIC_YingyanLin_Rice_DemingChen_UIUC_FPGA_2020_10.1145/3373087.3375306,amazon_ec2_f1, microsoft_brainwave,DNN_NAS_Hardware_YiyuShi_ICCAD_2019_Jiang:2019:AVE:3316781.3317757,DNN_NAS_StandingShoulder_YiyuShi_HardwareCoDesign_TCAD_2020_jiang2020standing,DNN_NAS_MultiFPGA_YiyuShi_Journal_TECS_2019_Jiang:2019:ASS:3365919.3358192}.

A common approach to \design is to use a nested loop:  the outer loop searches over the hardware space while the inner loop searches
for the optimal architecture given the hardware choice in the outer loop,
or vice versa (i.e., outer loop for architectures and
inner loops for hardware) \cite{yiyu_co_exploration,DNN_NAS_StandingShoulder_YiyuShi_HardwareCoDesign_TCAD_2020_jiang2020standing}.
Alternatively, one can also simultaneously search over the neural architecture and hardware
spaces as a combined design choice \cite{song_han_neural_accelerator_architecture_search}.

While \design can further optimize DNN performance 
\cite{yiyu_codesign_dac},
it also exponentially enlarges the search space, presenting significant challenges.
For example, the combination of architecture and accelerator design spaces
can be up to $10^{861}$ \cite{song_han_neural_accelerator_architecture_search}.
Concretely,
letting $M$ and $N$ be the sizes  of the architecture space
and hardware/accelerator space, respectively,
the
total search complexity is in the order of $\mathcal{O}(MN)$.
By contrast, the fully-decoupled approach (i.e., separately
performing NAS and accelerator optimization)
has a total complexity of $\mathcal{O}(M+N)$, although
it only results in sub-optimal designs.

Consequently, many studies have been focusing on
speeding up the evaluation of
co-design choices (e.g., using
accuracy predictor and latency/energy simulation instead
of actual measurement \cite{DNN_NAS_APQ_JointSearch_ArchitecturePruningQuantization_SongHan_CVPR_2020_Wang_2020_CVPR,DNN_NAS_OnceForAll_SongHan_MIT_2020_ICLR,DNN_AutoDNNChip_FPGA_ASIC_YingyanLin_Rice_DemingChen_UIUC_FPGA_2020_10.1145/3373087.3375306,DNN_NAS_Hardware_YiyuShi_ICCAD_2019_Jiang:2019:AVE:3316781.3317757}), and/or improving the search efficiency
(e.g., reinforcement learning or evolutionary search to co-optimize
architecture and hardware \cite{song_han_neural_accelerator_architecture_search,DNN_NAS_Hardware_YiyuShi_ICCAD_2019_Jiang:2019:AVE:3316781.3317757,DNN_NAS_MultiFPGA_YiyuShi_Journal_TECS_2019_Jiang:2019:ASS:3365919.3358192}).
Nonetheless,
 due to the $\mathcal{O}(MN)$ search space,
the SOTA \design is still a time-consuming process, taking
up a few or even tens of GPU hours for each new deployment scenario (e.g., changing
the latency and/or energy constraints) \cite{song_han_neural_accelerator_architecture_search,DNN_NAS_StandingShoulder_YiyuShi_HardwareCoDesign_TCAD_2020_jiang2020standing}.

\textbf{Contributions.}
By settling in-between the fully-decoupled approach
and the fully-coupled co-design approach,
we propose a new \emph{semi}-decoupled approach
 to reduce
 the size of the total co-search space $\mathcal{O}(MN)$
 by orders of magnitude, yet without losing design optimality.
Our approach builds on
the \textit{latency and energy monotonicity} --- the  architectures'
ranking orders in terms of inference latency and energy consumption on different accelerators are highly correlated
--- and includes two stages.
In \textbf{Stage~1},
we randomly choose a sample accelerator (a.k.a.,
a proxy accelerator), and then run hardware-aware NAS for $K$
times to find a set $\mathcal{P}$ consisting of $K=|\mathcal{P}|$ optimal architectures
for this proxy. Clearly, compared to 
$M$ and $N$,
the size of $\mathcal{P}$ is orders-of-magnitude smaller (e.g.,
10-20 vs. $10^{18}$ \cite{darts}).
Then, in \textbf{Stage~2}, instead of the entire architecture space as in the SOTA
co-design,
we only jointly
search over the hardware space combined with the small set $\mathcal{P}$,
which significantly reduces the total search space.
Crucially, by latency and energy monotonicity, the set of optimal architectures
is (approximately) the same for all accelerator designs,
and hence selecting architectures out of $\mathcal{P}$
can still yield the optimal or very close-to-optimal architecture
design.

We validate our approach by conducting experiments on a state-of-the-art neural accelerator simulator MAESTRO \cite{maestro_micro2019}.
Our results confirm that strong latency and energy monotonicity
exist among different accelerator designs. More importantly, by using one candidate accelerator as the proxy and obtaining its small set of optimal architectures, we can reuse the same architecture set  for other accelerator candidates during
the hardware search stage.

\pdfoutput=1

\section{Problem Formulation}

We focus on the design
of a single neural architecture-accelerator pair.
The main goal is to maximize
the inference accuracy subject to a few design constraints
such as inference latency, energy, and area \cite{DNN_NAS_StandingShoulder_YiyuShi_HardwareCoDesign_TCAD_2020_jiang2020standing}.
Next, by denoting
the neural architecture and hardware
as $a$ and $h$, respectively,
we formulate
 the problem 
as follows:
\begin{eqnarray}
\label{eqn:objective_original}
&\max_{{a}\in\mathcal{A},\; {h}\in\mathcal{H}}Accuracy({a})\\
\label{eqn:latency}
s.t.,&Latency({a},{h})\leq {L}\\
\label{eqn:energy}
&Energy({a},{h})\leq {E}\\
\label{eqn:hardware}
&HardwareResource(h)\leq H,
\end{eqnarray}
where the objective $Accuracy({a})$  depends on the architecture,\footnote{The inference accuracy also depends
on the network weight trained on a dataset, which is not a decision variable
in \design
and hence omitted.}
the first two constraints are
set on the inference latency and energy consumption that depend on both
 the architecture and hardware choices,
and the last constraint is on the hardware configuration itself (e.g., area) and hence
independent of the architecture.
We denote the optimal
design as $(a^*,h^*)$ which solves the optimization
problem Eqns.~\eqref{eqn:objective_original}---\eqref{eqn:hardware}.
Note that, because of the combinatorial nature of the problem,
\emph{optimality} is not in a mathematically strict sense;
instead, a design $(a,h)$
is often considered as \emph{optimal} if it is good enough
in practice (e.g., better
than or competitive
with SOTA designs).

Suppose that the architecture space $\mathcal{A}$ and hardware space
$\mathcal{H}$ have $M=|\mathcal{A}|$ and $N=|\mathcal{H}|$ design choices, respectively,
which are both extremely large in practice.
Thus, the co-design space $\mathcal{A}\times \mathcal{H}$ has a total of $MN$
architecture-hardware combinations.
 This makes
exhaustive search virtually impossible
and adds significant challenges to
co-design over
the joint search space.

\textbf{Remark.} In our formulation,
the notation of neural ``architecture'' $a\in\mathcal{A}$
can also broadly include
 other applicable design factors for the DNN model
 (e.g., weight quantization).
Moreover, the hardware $h$ implicitly includes
the dataflow design, which is a downstream task
based on the architecture
and hardware choices.
In the following, 
we also interchangeably use ``accelerator'' and ``hardware''
to refer to the hardware-dataflow combination
 unless otherwise specified. Thus,
 with
different dataflows, the same hardware configuration
will be considered as different $h\in\mathcal{H}$.

\pdfoutput=1

\section{A Semi-Decoupled Approach}
\label{sec: approach}

In this section, we first review the existing architecture-accelerator
design approaches,
and then present our \ouralg approach.

\subsection{Overview of Existing Approaches}

\subsubsection{Fully decoupled approach.}
A straightforward approach is to separately optimize architectures
and accelerators in a siloed manner by decoupling NAS from accelerator design \cite{DNN_AutoDNNChip_FPGA_ASIC_YingyanLin_Rice_DemingChen_UIUC_FPGA_2020_10.1145/3373087.3375306,DNN_NAS_APQ_JointSearch_ArchitecturePruningQuantization_SongHan_CVPR_2020_Wang_2020_CVPR,DNN_NAS_ChamNet_Prediction_CVPR_2019_dai2019chamnet}:
first perform NAS to find \emph{one} optimal architecture $\tilde{a}\in\mathcal{A}$, and then
optimize the accelerator design for this particular architecture $\tilde{a}$;
or, alteratively, first optimize the accelerator $\tilde{h}\in\mathcal{H}$, and then perform
NAS to find the optimal architecture for this particular accelerator
$\tilde{h}$. This approach has a total complexity in the order of
$\mathcal{O}(M+N)$ where $M=|\mathcal{A}|$ and $N=|\mathcal{H}|$.
But, the drawback is also significant: it does not fully
exploit the flexibility of the co-design space
and,
as shown in several prior studies \cite{DNN_NAS_StandingShoulder_YiyuShi_HardwareCoDesign_TCAD_2020_jiang2020standing,song_han_neural_accelerator_architecture_search,DNN_NAS_Hardware_YiyuShi_ICCAD_2019_Jiang:2019:AVE:3316781.3317757},
 can result
in highly sub-optimal architecture-accelerator
designs.

\subsubsection{Fully coupled approach.}
As can be seen in Eqns.~\eqref{eqn:latency} and~\eqref{eqn:energy},
the inference latency and energy consumption is jointly
determined by the architecture and hardware choices. Such entanglement of
architecture and hardware is the key reason for the SOTA \design.

Concretely,
a general co-design approach is to use a nested loop
\cite{DNN_NAS_Hardware_YiyuShi_ICCAD_2019_Jiang:2019:AVE:3316781.3317757}.
For example, the outer loop searches over the hardware space, whereas the inner loop searches
for the optimal architecture given the hardware choice in the outer loop.
Alternatively, another equivalent approach is to  first search for neural architectures
in the outer loop and then search for accelerators in the inner loop.

Here, we use ``outer loop for hardware and inner loop for architecture''
as an example.
While the actual search method can differ from one study to another
(e.g., reinforcement learning vs. evolutionary search \cite{DNN_NAS_Hardware_YiyuShi_ICCAD_2019_Jiang:2019:AVE:3316781.3317757,DNN_NAS_OnceForAll_SongHan_MIT_2020_ICLR}), this nested search can be mathematically formulated as a bi-level optimization problem
below:
\begin{eqnarray}
\label{eqn:objective_new}
\textbf{Outer:}&\max_{{h}\in\mathcal{H}} Accuracy({a}^*(h))\\
\label{eqn:hardware_new}
s.t.,&HardwareResource(h)\leq H,
\end{eqnarray}
where, given a choice of $h$,
the architecture ${a}^*(h)=a^*(h; L, E)$
solves the {inner} hardware-aware NAS problem:
\begin{eqnarray}
\label{eqn:objective_new_inner}
\textbf{Inner:}&\max_{{a}\in\mathcal{A}}Accuracy({a})\\
\label{eqn:latency_new}
s.t.,& Latency({a},{h})\leq {L}\\
\label{eqn:energy_new}
&Energy({a},{h})\leq {E}.
\end{eqnarray}
In Eqn.~\eqref{eqn:objective_new}, ${Accuracy(\cdot)}$
is still decided by the  architecture, although
we use ${a}^*(h)=a^*(h; L, E)$ to emphasize that
the architecture is specifically optimized for the given hardware candidate $h$.

We see that, during the search for the optimal
hardware $h^*$ in the outer problem,
the inner NAS problem is repeatedly solved as a subroutine and
yields the optimal architecture ${a}^*(h)=a^*(h; L, E)$ given each hardware
choice $h$ set by the outer search. For notational convenience,
we also use ${a}^*(h)$ to represent $a^*(h; L, E)$ without causing ambiguity.

The focus of SOTA \design
approaches have been primarily on speeding up the evaluation of
architecture-hardware choices (e.g., using
accuracy predictor and latency/energy simulation instead
of actual measurement \cite{DNN_NAS_APQ_JointSearch_ArchitecturePruningQuantization_SongHan_CVPR_2020_Wang_2020_CVPR,DNN_NAS_OnceForAll_SongHan_MIT_2020_ICLR,DNN_AutoDNNChip_FPGA_ASIC_YingyanLin_Rice_DemingChen_UIUC_FPGA_2020_10.1145/3373087.3375306,DNN_NAS_Hardware_YiyuShi_ICCAD_2019_Jiang:2019:AVE:3316781.3317757}), and/or improving the search efficiency
(e.g., reinforcement learning or evolutionary search to co-optimize
architecture and hardware \cite{song_han_neural_accelerator_architecture_search,DNN_NAS_Hardware_YiyuShi_ICCAD_2019_Jiang:2019:AVE:3316781.3317757,DNN_NAS_MultiFPGA_YiyuShi_Journal_TECS_2019_Jiang:2019:ASS:3365919.3358192}).
 Nonetheless,
evaluating one architecture-accelerator combination
can still take up a few seconds in total (e.g., running MAESTRO to
perform mapping/scheduling
and estimate the latency and energy
consumption takes 2-5 seconds on average \cite{maestro_micro2019}).
Then, compounded by the exponentially large architecture and hardware space
in the order of $\mathcal{O}(MN)$,
the total \design cost is very high (e.g.,
a few or even tens of GPU hours for
each deployment scenario \cite{song_han_neural_accelerator_architecture_search,DNN_NAS_Hardware_YiyuShi_ICCAD_2019_Jiang:2019:AVE:3316781.3317757}).

\begin{table}[!t]
\caption{Comparison of Different Approaches}
\vspace{-0.3cm}
\small
\begin{tabular}{|l|c|c|}
\hline
{\textbf{Approach}} & {\textbf{Optimality}} & {\textbf{Complexity}} \\ \hline
Fully-decoupled separate design & No & $\mathcal{O}(M+N)$  \\
Fullly-coupled co-design & Yes & $\mathcal{O}(MN)$  \\
\textbf{Semi-decoupled co-design} & Yes & $\mathcal{O}(K(M+N))$  \\
\hline
\end{tabular}
\label{tab:compare}
\end{table}

\subsection{Semi-Decoupled Co-Design}

We propose a 
\ouralg approach ---
partially decoupling NAS from hardware search
to reduce the total co-search cost from $\mathcal{O}(MN)$
to $\mathcal{O}(K(M+N))$ in a principled manner,
where $K$ is orders-of-magnitude less than
$M$ and $N$.

\textbf{Performance monotonicity.}
The key intuition underlying our \ouralg
approach is the latency and energy performance monotonicity ---
given different accelerators,
the architectures' ranking orders in terms of both the inference latency and energy consumption are highly correlated.
We can measure the ranking correlation
in terms of the Spearman's rank correlation coefficient (SRCC),
whose value lies within $[-1,1]$ with ``$1$'' representing
the identical ranking orders \cite{srcc_guide}.

It has been shown in a recent hardware-aware NAS study
\cite{Shaolei_NAS_OneProxy_Sigmetrics_Journal_Dec2021}
that the architectures' ranking orders in terms of inference latency
 are highly similar on different devices, with SRCCs often close to 0.9 or higher, especially among
 devices of the same platform (e.g., mobile phones).
 For example,
 if one architecture $a_1$ is faster than
another architecture $a_2$ on one mobile phone, then it is very likely that
$a_1$ is still faster than $a_2$ on another phone.
One reason is
that architectures are typically either computing-bound
or memory-bound on devices of the same platform, which, by roofline analysis, results in similar rankings of their latencies \cite{Roofline_Patterson_CACM_2009_10.1145/1498765.1498785}.
 Based on this property (a.k.a., latency monotonicity), it has been theoretically
and empirically proved
that the Pareto-optimal architectures on different devices are highly
overlapping if not identical \cite{Shaolei_NAS_OneProxy_Sigmetrics_Journal_Dec2021}.

While the target hardware space chosen
by the designer has many
choices, it essentially covers
one platform --- neural accelerator under a set of hardware constraints.
As a result, we expect latency monotonicity to be satisfied in our problem.
Additionally, beyond the findings in \cite{Shaolei_NAS_OneProxy_Sigmetrics_Journal_Dec2021},
we observe in our experiments that \emph{energy} monotonicity also holds:
if one architecture $a_1$ is more energy-efficient than
another architecture $a_2$ for one hardware choice, then it is very likely that
$a_1$ is still more energy-efficient than $a_2$ for another hardware choice.
Along with latency monotonicity, energy monotonicity will
be later validated in our experiments.
 One reason for the energy monotonicity is that energy consumption is highly
 related to the inference latency with a strong correlation \cite{DNN_NAS_HardwareNAS201_Benchmark_Rice_ICLR_2021_li2021hwnasbench}.

 For simplicity, we use \emph{performance} monotonicity to collectively refer to
both latency and energy monotonicity.

\textbf{Insights.} The performance monotonicity
leads to the following proposition, which generalizes
the statement in \cite{Shaolei_NAS_OneProxy_Sigmetrics_Journal_Dec2021}
by considering both latency and energy monotonicity.
We first note that, by solving the inner NAS problem
under a set of latency and energy constraints
in Eqns.~\eqref{eqn:objective_new_inner}---\eqref{eqn:energy_new},
we can construct a set
$\mathcal{P}(h)=(a_{1}^*(h;L_1,E_1),\cdots, a_K^*(h; L_K, E_K))$
of optimal architectures covering the architectures along
the Pareto boundary. The size $K=|\mathcal{P}(h)|$ of the optimal architecture
set depends on the granularity
of latency and energy constraints we choose. In practice,
$K$ in the order of a few tens (e.g., $10-30$) is sufficient to cover a
wide range of latency and energy constraints for our design target.

\begin{proposition}\label{prop:pareto}
Given performance monotonicity,
the set of optimal architectures
$\mathcal{P}(h)=(a_{1}^*(h;L_1,E_1),\cdots, a_K^*(h; L_K, E_K))$
found by the inner hardware-aware NAS problem in Eqns.~\eqref{eqn:objective_new_inner}---\eqref{eqn:energy_new}
is the same for all hardware choices, i.e.,
$\mathcal{P}(h_1) = \mathcal{P}(h_2)$, for all
 $h_1,h_2\in\mathcal{H}$.
 \proof Consider two
 hardware choices $h_1,h_2\in\mathcal{H}$.
 By performance monotonicity, we can replace the constraints
 $Latency({a},{h_1})\leq {L_1}$ and $Energy({a},{h_2})\leq {E_1}$
 with another two equivalent constraints
 $Latency({a},{h_2})\leq {L_1}'$ and $Energy({a},{h_2})\leq {E_1}'$, respectively.
By varying ${E_1}$ and $L_1$ over their feasible ranges,
we obtain the optimal architecture set
$\mathcal{P}(h_1)$ for $h_1$.
Accordingly, due
to the equivalent latency and energy constraints for $h_2$,
we also obtain the optimal architecture set
$\mathcal{P}(h_2)$ for $h_2$, thus completing
the proof. $\hfill\blacksquare$
\end{proposition}

\begin{figure}[!t]
		\centering \includegraphics[trim=0 4.5cm 9.5cm 0, clip, width=0.4\textwidth]{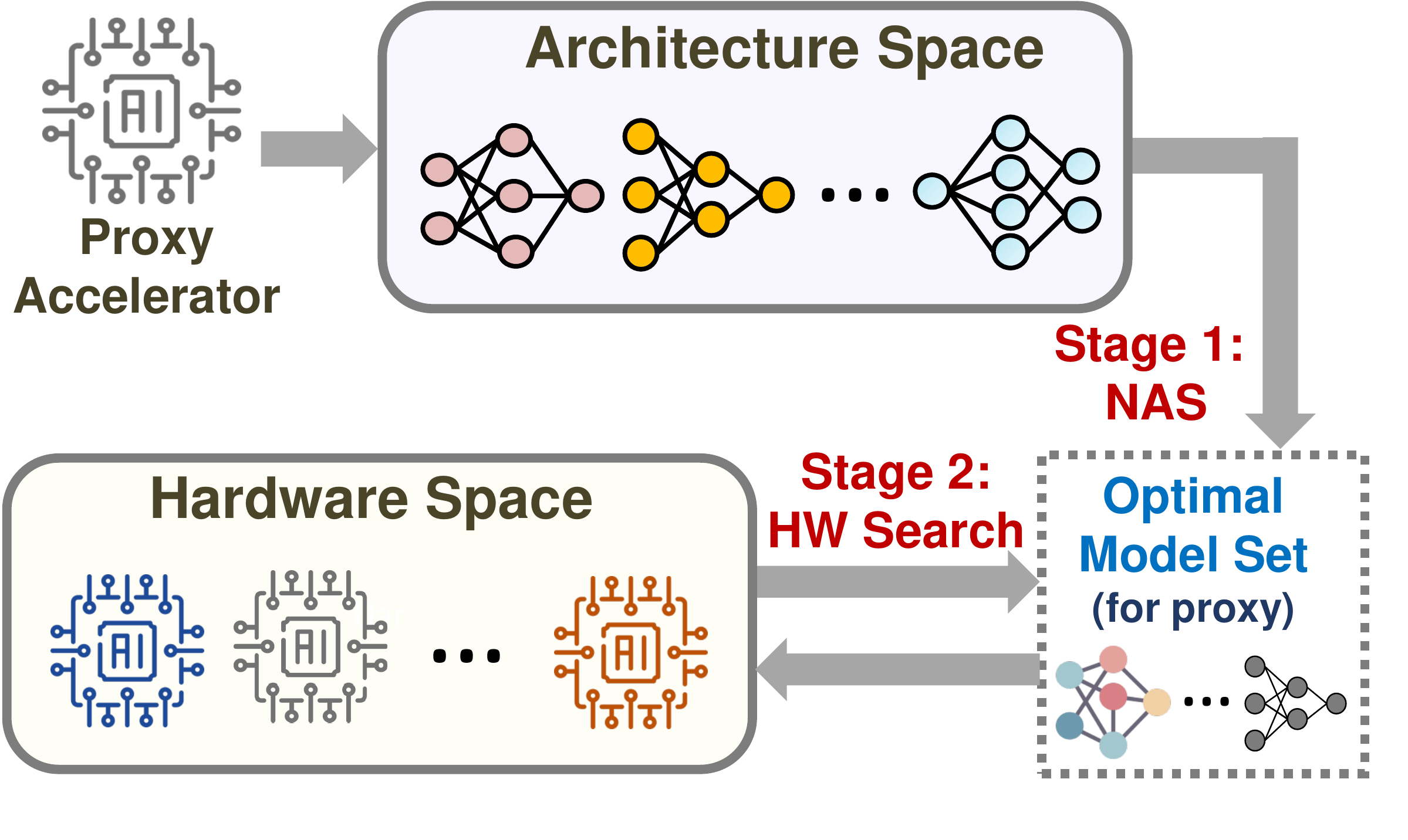}	
	\vspace{-0.5cm}
	\caption{Overview of our
\ouralg approach.} 
	\label{fig: framework_new}
\end{figure}

Proposition~\ref{prop:pareto} ensures that in the presence
of performance monotonicity, the same set $\mathcal{P}(h)$
of optimal architectures apply to all $h\in\mathcal{H}$. Thus, we can also
simply
use $\mathcal{P}$ to denote the set of optimal architectures,
which are essentially \emph{shared} by $h\in\mathcal{H}$.
Note carefully that Proposition~\ref{prop:pareto} does \emph{not} mean that, given a specific pair of latency and energy constraints, we will have
the same architecture $a^*(h_1; L, E)=a^*(h_2; L, E)$ for two hardware choices $h_1,h_2\in\mathcal{H}$.

Nonetheless, once we have found $\mathcal{P}\subset \mathcal{A}$,
there is no need to jointly search over the entire architecture-hardware space
$\mathcal{A}\times \mathcal{H}$
any more. Instead, it is sufficient to merely search over the
restricted architecture-hardware space  $\mathcal{P}\times \mathcal{H}$.
Importantly, the set $\mathcal{P}$ of optimal arachitectures
is orders-of-magnitude smaller than the entire architecture space
$\mathcal{A}$ (e.g., a few tens vs. $10^{18}$ in the DARTS architecture space \cite{darts}), thus significantly reducing the total \design cost without
losing optimality.

\begin{algorithm}[!t]
	\caption{Semi-Decoupled Architecture-Accelerator Co-Design}\label{algo:new_algorithm}
	\begin{algorithmic}[1]
		\STATE $\textbf{Input:}$ Architecture space $\mathcal{A}$,
hardware space $\mathcal{H}$, sample hardware $h_0\in\mathcal{H}$, and design constraints
$L,E,H$ in Eqns.~\eqref{eqn:latency},~\eqref{eqn:energy},~\eqref{eqn:hardware}
		\STATE $\textbf{Output:}$ Optimal co-design $(a^*,h^*)$
		\STATE	$\textbf{Initilization:}$ \texttt{Choose $K$ latency and energy constraints $(L_k,E_k)$ for $k=1,\cdots,K$, set $\mathcal{P}=\varnothing$,
and randomly choose $(a^*,h^*)$};
        \FOR{\texttt{$k=1,\cdots, K$}}
		\STATE \texttt{For constraints $(L_k,E_k)$, run hardware-aware NAS to get
optimal architecture $a_k^*(h_0; L_k, E_k)$}
		\STATE \texttt{$\mathcal{P}=\mathcal{P}\bigcup \{a_k^*(h_0; L_k, E_k)\}$;}
		\ENDFOR
		\FOR{\texttt{each candidate hardware $h\in\mathcal{H}$}}
		\IF{\texttt{$HardwareResource(h)\leq H$}}
        \STATE \texttt{Find optimal architecture $a^*(h)\in\mathcal{P}$ satisfying
the latency and energy constraint $(L,E)$}
        \IF{\texttt{$Accuracy(a^*(h))>Accuracy(a^*)$}}
        \STATE \texttt{$(a^*,h^*)\leftarrow(a^*(h),h)$}
        \ENDIF
        \ENDIF
		\ENDFOR
	\end{algorithmic}
\end{algorithm}

\textbf{Algorithm.}
Our \ouralg approach has two stages,
as illustrated in Fig.~\ref{fig: framework_new}
and summarized in Algorithm~\ref{algo:new_algorithm}.

\textbf{Stage 1:} We randomly choose a sample accelerator $h_0\in\mathcal{H}$, which we refer to as
the \emph{proxy} accelerator, and run hardware-aware NAS for
$K$ times to find a
set of optimal architectures
$\mathcal{P}=\mathcal{P}(h_0)=(a_{1}^*(h_0;L_1,E_1),\cdots, a_K^*(h_0; L_K, E_K))$.
Specifically, 
 $\mathcal{P}$ is constructed by setting
$K$ different latency and energy constraints and accordingly solving the inner NAS
problem in Eqns.~\eqref{eqn:objective_new_inner}---\eqref{eqn:energy_new}
for $K$ times. Thus, the search cost in Stage 1 is $\mathcal{O}(KM)$ where
$M=|\mathcal{A}|$.

\textbf{Stage 2:} We search for the optimal accelerator
$h^*\in\mathcal{H}$.
Specifically, given each candidate $h\in\mathcal{H}$
(selected by, e.g., reinforcement learning or evolutionary
search \cite{DNN_NAS_Hardware_YiyuShi_ICCAD_2019_Jiang:2019:AVE:3316781.3317757,song_han_neural_accelerator_architecture_search}),
instead of searching over the entire architecture set $\mathcal{A}$,
we obtain its corresponding optimal architecture $a^*(h)$ from the
set $\mathcal{P}\subset \mathcal{A}$ constructed in Stage 1.
Thus, the search cost in Stage 2 is $\mathcal{O}(KN)$ where
$N=|\mathcal{H}|$.

\subsection{Discussion}

In practice, 
performance monotonicity
may not be perfectly satisfied.
Thus, the optimal architecture $a^*(h)$ corresponding
to a candidate accelerator $h\in\mathcal{H}$
may not always strictly belong to the optimal architecture set
$\mathcal{P}$ that is pre-constructed based on the proxy $h_0$.
Nonetheless, by only searching over $\mathcal{P}$ for this candidate accelerator $h$,
we can still find an architecture
$a\in\mathcal{P}$ that is
\emph{close}-to-optimal.
In fact,
to speed up the NAS process and find competitive architectures,
 it is
very common to use proxy/substitute metrics
(such
as accuracy predictor or the neural tangent kernel \cite{DNN_NAS_NTK_Proxy_ZhangyangWang_UTAustin_ICLR_2021_chen2021neural})
which only have SRCC
of around 0.5--0.9 with the true performance.
In our problem, we can also view the architectures' latency and energy
performance on the proxy accelerator $h_0$ as the substitute performance
on other accelerator candidates.
Therefore, given the good
albeit not necessarily close-to-perfect
performance monotonicity,
 the architectures
optimized specifically for the proxy
are also
 sufficiently competitive ones
 for other accelerator candidates.

In \cite{Shaolei_NAS_OneProxy_Sigmetrics_Journal_Dec2021},
scalable hardware-aware NAS is proposed by utilizing
latency monotonicity on various devices.
Without considering energy consumption, a
high SRCC (>0.9) for latency
is needed to ensure that one proxy device's optimal architectures
are still close to optimal on another device.
In our problem, such high SRCC values are not necessarily needed,
because we consider both energy and latency --- moderate SRCC
values on two performance metrics are enough. This is reflected
in both our experiments and prior studies (e.g.,
two proxy metrics having moderate SRCC values with the true accuracy
can estimate the accuracy performance very well \cite{DNN_NAS_NTK_Proxy_ZhangyangWang_UTAustin_ICLR_2021_chen2021neural}).

In the highly unlikely event of very low SRCCs
 (e.g., 0.2) between the proxy and other accelerator
candidates, we can enlarge $\mathcal{P}$
by adding some approximately optimal architectures near the Pareto boundary (for the chosen
proxy), such that they can be competitive choices
for other candidate accelerators.
Alternatively, we could use \emph{a few} proxy accelerators, each
having good latency and energy monotonicity with a subspace
of accelerator design, and jointly construct an expanded set
 $\mathcal{P}$ of optimal architectures in Stage~1.
In any case,
the set $\mathcal{P}$ is orders-of-magnitude smaller
than the entire architecture space or accelerator space.

\textbf{Summary.}
 The essence of our semi-decoupled approach is to
use a proxy $h_0$ to find a small set of optimal
architectures that also includes the actual optimal
or \emph{close}-to-optimal architectures for different
accelerator candidates, thus reducing the total co-design complexity
without losing optimality.
This is significantly different from a typical fully-decoupled approach that pre-searches
for \emph{one} architecture and then find the matching accelerator,
and also has a sharp contrast with a fully-coupled co-design approach that
jointly searches over the entire architecture-accelerator space.
The comparison of different approaches is also summarized in Table~\ref{tab:compare}.
Importantly,
our approach focuses on reducing
the search space complexity, and can be integrated with any actual
NAS (Stage 1) and accelerator exploration techniques
(Stage 2).

\pdfoutput=1

\section{Experiment Setup}
\label{sec: experiment_setup}

We provide details of our experiment setup as follows.

\textbf{Accelerator hardware space.}
We employ an open-source tool MAESTRO \cite{maestro_micro2019} to simulate DNNs on the accelerator and measure inference metrics (e.g., latency and energy).
MAESTRO supports a wide range of accelerators, including global shared scratchpad (i.e., L2 scratchapd), local PE scratchpad (i.e., L1 scratchpad), NoC, and a PE array organized into different 
hierarchies or dimensions.

\textbf{DNN dataflow.}
Dataflow decides the DNN partitioning and scheduling strategies, which affects inference latency and energy performance.
We consider three template dataflows: \textbf{KC-P} (motivated by NVDLA~\cite{NVDLA_Deep_Learning_Accelerator}), \textbf{YR-P} (motivated by Eyeriss~\cite{eyeriss}), and \textbf{X-P} (weight-stationary).
Exhibiting different characteristics (e.g., temporal reuse
of input activation and filter in YR-P vs. spatial reuse of
input activation in KC-P), these representative dataflows are all supported
by MAESTRO \cite{maestro_micro2019} and commonly used
in SOTA \design \cite{yiyu_codesign_dac}.

\textbf{Architecture space.} We consider the following two spaces.

$\bullet$ \emph{NAS-Bench-301}: It is a SOTA surrogate NAS benchmark built via deep ensembles and modeling uncertainty, which provides close-to-real predicted performances (i.e., accuracy and training time) of $10^{18}$ architectures on CIFAR-10 \cite{DNN_NAS_Benchmark_SurrogateBenchmarks_301_2020_siems2020nasbench301}.
We consider the DARTS space \cite{darts}, where each architecture is a stack of 20 convolutional cells, and each cell consists of seven nodes.

$\bullet$ \emph{AlphaNet}: It
is a new family of architectures on ImageNet discovered by applying a generalized $\alpha$-divergence to supernet training \cite{alphanet_facebook_2021}.
Our search space is based on Table~7 of \cite{alphanet_facebook_2021}, with a slight variation that the channel width is fixed as "16, 16, 24, 32, 64, 112, 192, 216, 1792", and depth, kernel size, expansion ratio of the first and last inverted residual blocks are fixed as "1, 1", "3, 3", "1, 6", respectively. For other searchable inverted residual blocks, the candidate depth, kernel size, and expansion ratio are "2, 3, 4, 5, 6", "3, 5, 7", and "3, 4, 6", respectively.

\textbf{Search strategy.} 
Our approach can be integrated
with any NAS and hardware search strategies. Here,
we consider exhaustive search over a pre-sampled subspace.
Specifically, for the NAS-Bench-301, we first
sample 10k models. Then,
 based on the accuracy given by NAS-Bench-301 and FLOPs of these 10k models, we select 1017 models, including the Pareto-optimal front (in terms of predicted accuracy and FLOPs)
 and some random architectures. Similarly, for the AlphaNet space,
 we first sample 10k models and then select 1046 models based on the predicted accuracy given by the released accuracy predictor \cite{alphanet_github} and FLOPs.
We consider a filtered space of 1k+ architectures (which include
the Pareto-optimal ones out of the 10k sampled architectures), because
using MAESTRO to measure the latency and energy of 10k models on thousands of different
hardware-dataflow combinations is beyond our computational resource limit.
For each of the three template dataflows,
 we sample 51 neural accelerators with different number of PEs, NoC bandwidth, and off-chip bandwidth per the MAESTRO document \cite{maestro_document}.
Specifically, the number of PEs can be chosen from ``512, 256, 128, 64, 32, 16", candidate NoC bandwidths are from ``300, 400, 500, 600, 700, 800, 900, 1000", and off-chip bandwidths are from ``50, 100, 150, 200, 250, 275, 300, 325, 350".
Note that
some of our sampled hardware-dataflow pairs
are not supported when running with KC-P and YR-P dataflows on MAESTRO.
Thus, the actual numbers of sampled accelerators
(i.e., hardware-dataflow combinations) are 133
for NAS-Bench-301 and 132 for AlphaNet, respectively.
We also consider layer-wise mixture of different dataflows (Section~\ref{sec:experiment_mixing}) to create 5000 different hardware-dataflow
combinations.

\pdfoutput=1

\section{Experimental Results}
\label{sec: experiment_results}

In this section, we present our experimental results.
We show
that strong performance monotonicity exists in the hardware design space,
and highlight that 
our \ouralg approach can
identify the optimal design at a much lower search
complexity.

\subsection{NAS-Bench-301}

\subsubsection{Performance monotonicity}
\begin{figure}[!t]\centering
	\subfigure[Latency SRCC]{
		\centering \includegraphics[width=0.295\linewidth]{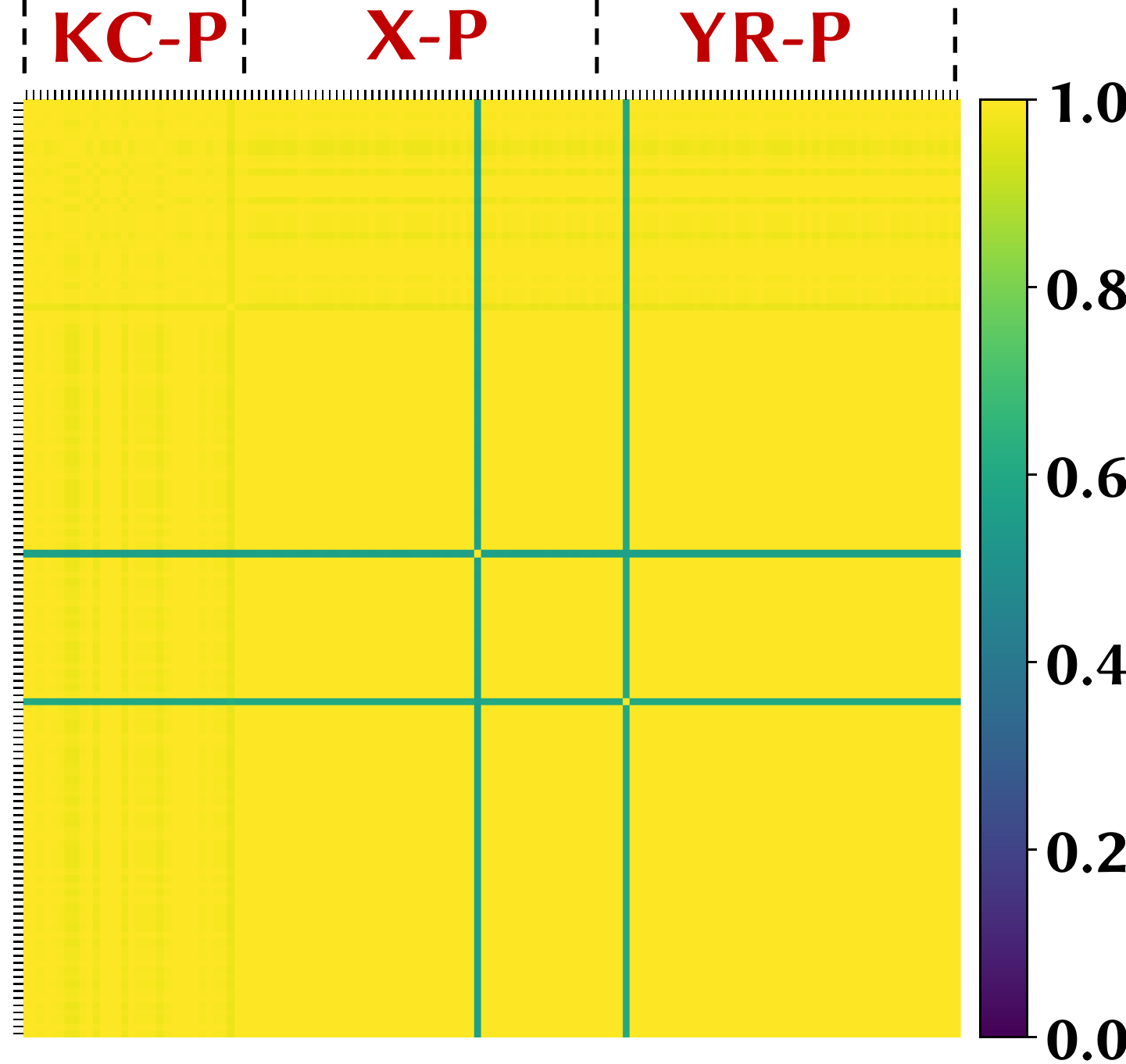}
		\label{fig: nasbench_latency}
	}
	\subfigure[Energy SRCC]{
		\centering
		\includegraphics[width=0.295\linewidth]{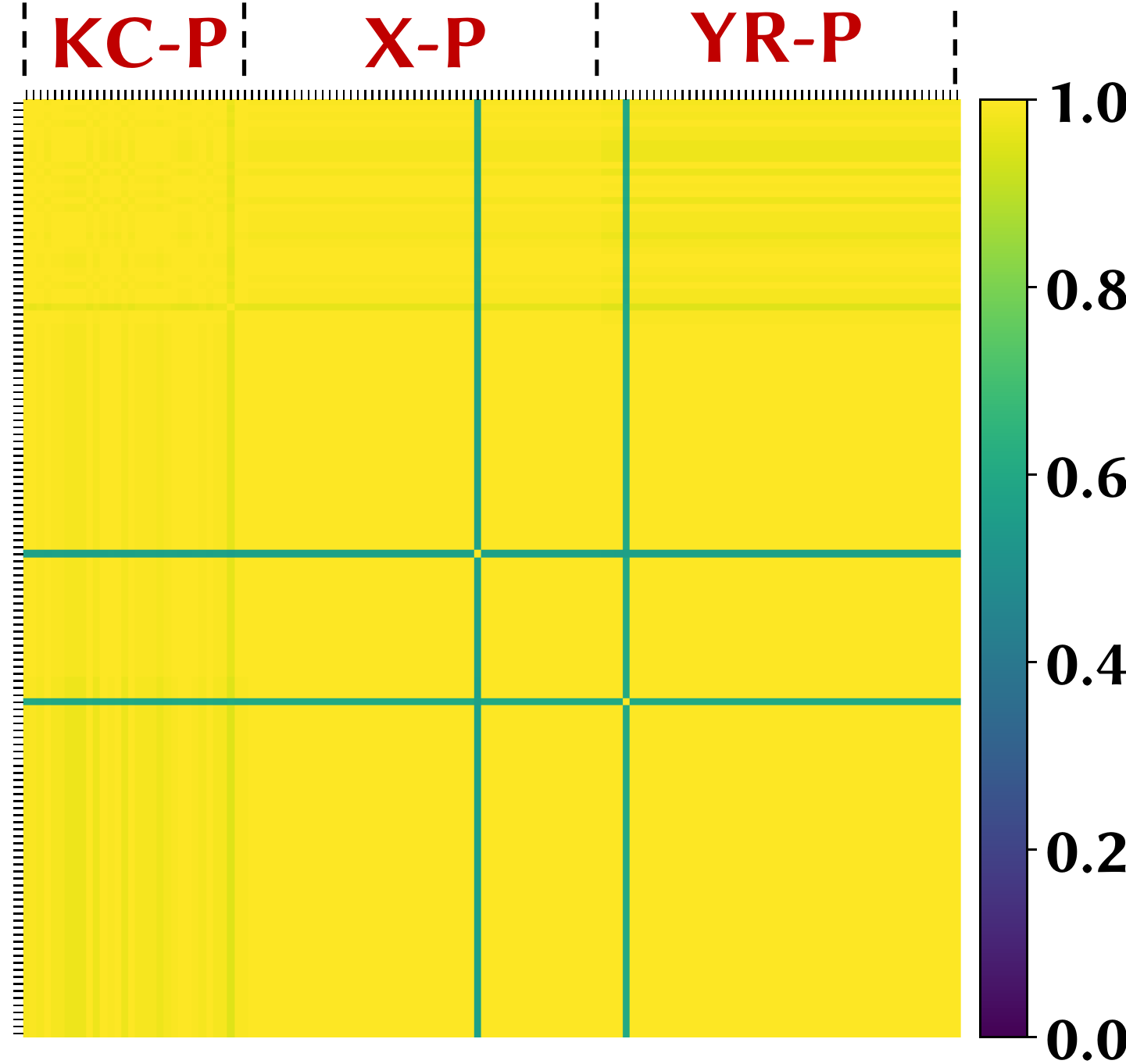}
		\label{fig: nasbench_energy}
	}
	\subfigure[CDF of SRCC]{
		\centering
\includegraphics[trim=0 0 0 0, clip, width=0.33\linewidth]{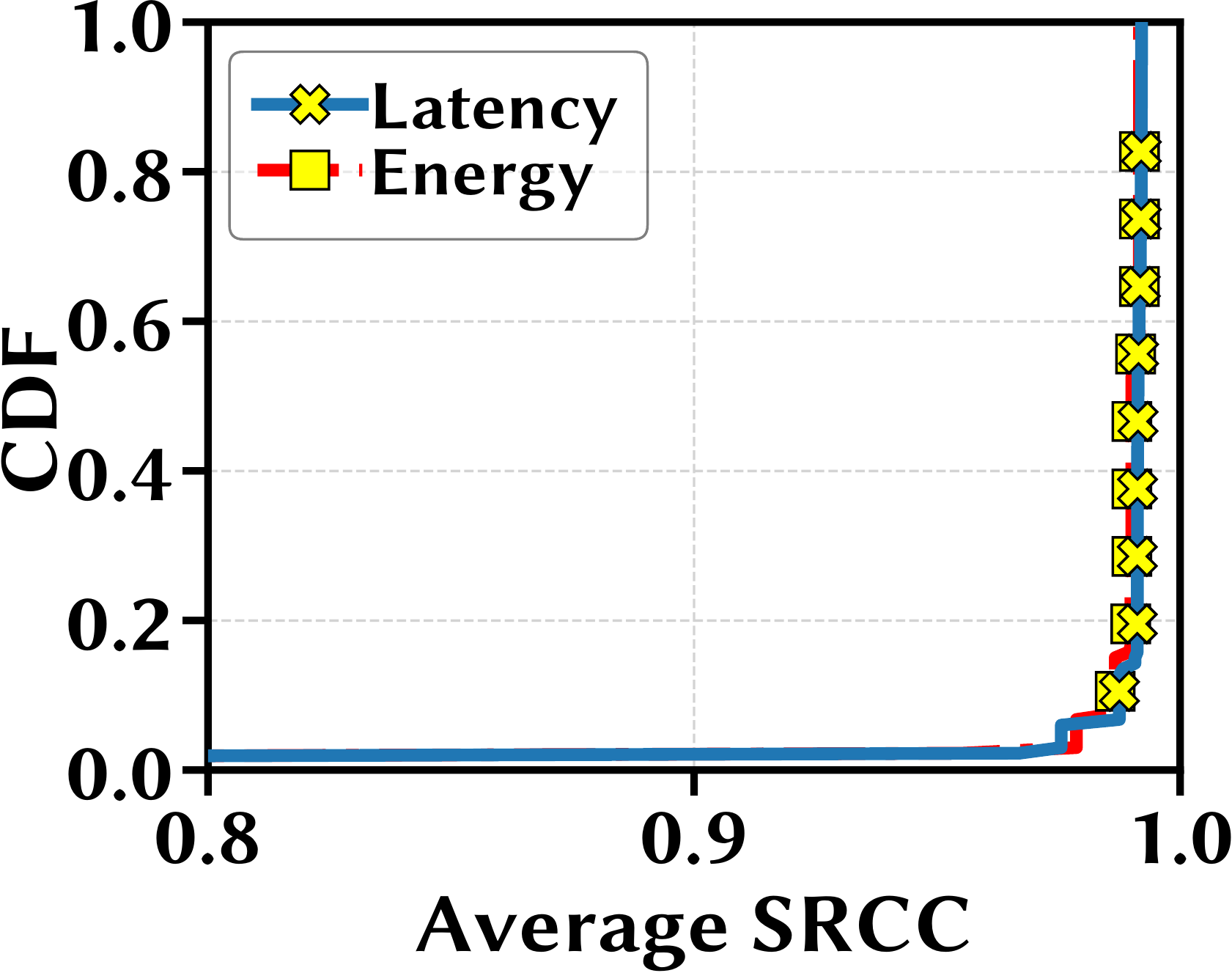}
		\label{fig: nasbench_cdf}
	}
	\vspace{-0.4cm}
	\caption{Performance monotonicity. We test 1017 models sampled in DARTS search space on 133 accelerators.}
	\vspace{-0.3cm}
	\label{fig: nasbench_srcc}
\end{figure}

We first validate that strong latency and energy performance monotonicity,
quantified in SRCC, holds between different accelerators.
The results are shown in Fig.~\ref{fig: nasbench_srcc}. We see
that,  except for two accelerator choices that have SRCC less than 0.6 with others, all the other accelerators have almost perfect performance monotonicity with SRCC greater
than 0.97. We also plot in Fig.~\ref{fig: nasbench_cdf}
the  cumulative distribution function (CDF) of the average SRCC values
for all the sampled accelerators, where for each accelerator $h$
the ``average'' is over the SRCC
values of all the accelerator pairs that include $h$.
We see that the vast majority of the accelerators have average SRCC
close to 1.

\begin{figure}[!t]
	\centering
	\includegraphics[width=0.48\textwidth]{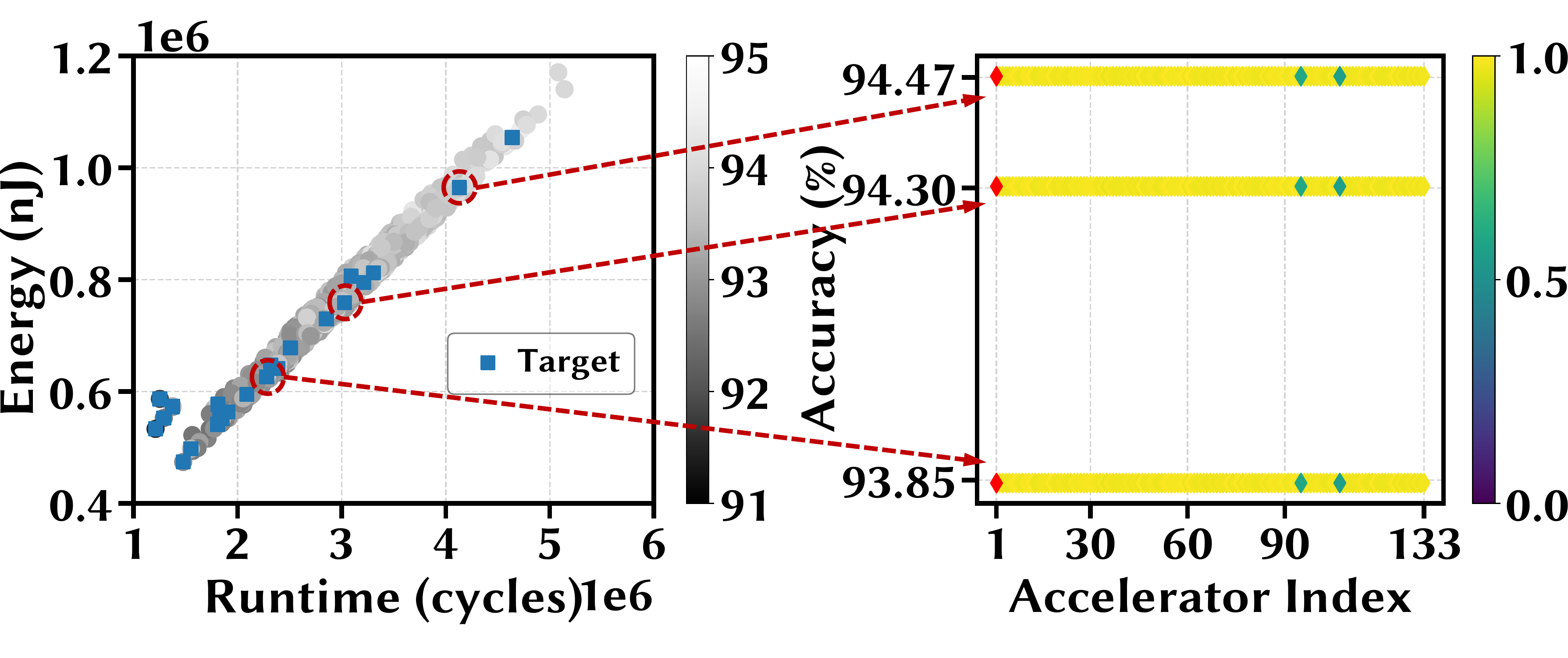}
\vspace{-0.6cm}
	\caption{NAS-Bench-301. Left: The optimal
models are marked in blue, and the grey scale indicates accuracy.
Right: The  accuracy of the model selected
from the proxy's optimal model set. We test each accelerator
as a different proxy.
\note{We also select two proxy accelerators (indexes 95 and 107)
that have the lowest SRCCs with the target,
and show the detailed results in Table~\ref{table: nasbench_pareto_3_config1}.}
}
\vspace{-0.3cm}
	\label{fig: nasbench_pareto_3}
\end{figure}

\begin{table*}[!t]\normalsize
	\centering
	\begin{adjustbox}{width=0.9\textwidth}
		\begin{tabular}{ccccccccccc}
			\hline
			{\multirow{2}*{\textbf{\tabincell{c}{Accelerator \\ Index}}}}
			& \multicolumn{2}{c}{\textbf{SRCC}}
			
			&  \multicolumn{4}{c}{\textbf{Hardware Config.}} & \multicolumn{3}{c}{\textbf{Model Performance}}  \\
			& \textbf{Latency} & \textbf{Energy}  &  \textbf{PEs}  & \textbf{NoC}  & \textbf{Off-chip}  & \textbf{Dataflow}  & \textbf{Latency (cycles)}  & \textbf{Energy (nJ)} & \textbf{Accuracy (\%)}\\
			\hline
			\hline
			1 (\textbf{target}) & 1 & 1 & 512 &  900 & 350 & KC-P & 2279256 & 626090 & 93.85 \\
			\cdashline{1-10}[0.8pt/2pt]
			107 & 0.556 & 0.567 &  64 &  400 & 250 & YR-P &   2279256 & 626090 & 93.85 \\
			\cdashline{1-10}[0.8pt/2pt]
			95  & 0.595 & 0.595 &  256 &  800 & 350 & X-P & 2279256 & 626090 & 93.85 \\
			\hline
			1 (\textbf{target}) & 1 & 1 & 512 &  900 & 350 & KC-P &   3027992 & 758928 & 94.30\\
			\cdashline{1-10}[0.8pt/2pt]
			107 & 0.556 & 0.567 & 64 &  400 & 250 & YR-P & 3027992 & 758928 & 94.30\\
			\cdashline{1-10}[0.8pt/2pt]
			95  & 0.595 & 0.595 &  256 &  800 & 350 & X-P &   3027992 & 758928 & 94.30\\
			\hline
			1 (\textbf{target}) & 1 & 1 & 512 &  900 & 350 & KC-P &  4130699 & 964783 & 94.47\\
			\cdashline{1-10}[0.8pt/2pt]
			107 & 0.556 & 0.567 &  64 &  400 & 250 & YR-P & 4130699 & 964783 & 94.47\\
			\cdashline{1-10}[0.8pt/2pt]
			95  & 0.595 & 0.595 &  256 &  800 & 350 & X-P &  4130699 & 964783 & 94.47\\
			\hline
		\end{tabular}
	\end{adjustbox}
	\vspace{0.2cm}
	\caption{\note{Hardware configuration of the target and two proxy accelerators, and performance metrics of the selected optimal models on each of them.
	``Accelerator Index" corresponds to the $x$-axis in right of Fig.~\ref{fig: nasbench_pareto_3}, the models on the target accelerator correspond to the circled ones in left of Fig.~\ref{fig: nasbench_pareto_3}, while the models on the two proxy accelerators correspond to the diamond marks located on the accelerator indexes. The architecture configuration of the target models is further illustrated in Table~\ref{table: nasbench_pareto_3_config2}.}}
	\label{table: nasbench_pareto_3_config1}
\end{table*}

\begin{table*}[!t]\normalsize
	\centering
	\begin{adjustbox}{width=0.9\textwidth}
		\begin{tabular}{ccccc}
			\hline
			{\multirow{2}*{\textbf{\tabincell{c}{Target Model}}}}  &
			\multicolumn{4}{c}{\textbf{Model Architecture}}  \\
			 & \textbf{Normal Cell Config.} & \textbf{Normal Cell Concat.}  &  \textbf{Reduce Cell Config.}  & \textbf{Reduce Cell Concat.} \\
			\hline
			\hline
			\#1 & \tabincell{c}{(skip\_connect, 0), (skip\_connect, 1), \\ (skip\_connect, 0), (skip\_connect, 2), \\ (sep\_conv\_5x5, 0), (skip\_connect, 1), \\ (dil\_conv\_5x5, 4), (skip\_connect, 2)} & [2, 3, 4, 5] & \tabincell{c}{(sep\_conv\_3x3, 1), (sep\_conv\_3x3, 0), \\ (dil\_conv\_3x3, 2), (skip\_connect, 0), \\ (sep\_conv\_5x5, 2), (avg\_pool\_3x3, 0), \\ (dil\_conv\_3x3, 3), (sep\_conv\_3x3, 1)} & [2, 3, 4, 5] \\
			\cdashline{1-5}[0.8pt/2pt]
			
			\#2 & \tabincell{c}{(skip\_connect, 0), (max\_pool\_3x3, 1), \\ (sep\_conv\_3x3, 0), (skip\_connect, 1), \\ (skip\_connect, 0), ('sep\_conv\_5x5, 3), \\ (avg\_pool\_3x3, 4),  (sep\_conv\_5x5, 1)} & [2, 3, 4, 5] & \tabincell{c}{(sep\_conv\_3x3, 1), (sep\_conv\_5x5, 0), \\ (avg\_pool\_3x3, 0), (sep\_conv\_5x5, 1), \\ (dil\_conv\_5x5, 3), (sep\_conv\_3x3, 2), \\ (avg\_pool\_3x3, 4), (sep\_conv\_3x3, 0)} & [2, 3, 4, 5] \\
			\cdashline{1-5}[0.8pt/2pt]
			
			\#3 & \tabincell{c}{(dil\_conv\_5x5, 0), (skip\_connect, 1), \\ (max\_pool\_3x3, 0), (max\_pool\_3x3, 2), \\ (sep\_conv\_5x5, 0),  (dil\_conv\_3x3, 3), \\ (dil\_conv\_5x5, 3), (dil\_conv\_5x5, 4)} & [2, 3, 4, 5] & \tabincell{c}{(skip\_connect, 0), (dil\_conv\_3x3, 1), \\ (sep\_conv\_3x3, 1), (sep\_conv\_5x5, 2), \\ (skip\_connect, 1), (max\_pool\_3x3, 0), \\ (skip\_connect, 1), (sep\_conv\_5x5, 2)} & [2, 3, 4, 5] \\
			\cdashline{1-5}[0.8pt/2pt]
			
			
			\hline
		\end{tabular}
	\end{adjustbox}
	\vspace{0.2cm}
	\caption{\note{Architecture configuration of the target models in Fig.~\ref{fig: nasbench_pareto_3}. The first row (i.e., target model \#1) corresponds to the leftmost circled model in Fig.~\ref{fig: nasbench_pareto_3}, and second row corresponds to the middle circled model, etc.
			These are the configurations for each convolutional cell constructing a complete model, which is a stack of 20 cells.
			For detailed explanation of the operations in the DARTS search space, please refer to \cite{darts} and \cite{DNN_NAS_Benchmark_SurrogateBenchmarks_301_2020_siems2020nasbench301}.}}
	\label{table: nasbench_pareto_3_config2}
\end{table*}

\subsubsection{Effectiveness}
To demonstrate the effectiveness,
 suppose that we have
an optimal architecture-accelerator pair $(a^*,h^*)$  produced
by the SOTA \design. We refer to the optimal accelerator
as the ``Target''.
By using our approach,
in Stage 1, we first randomly choose a non-target accelerator $h_0$ as our proxy,
and run hardware-aware NAS on this proxy to obtain
the set $\mathcal{P}$ of optimal architectures.
Next, in Stage 2, we will search over the accelerator space, retrieve
the corresponding architecture $a_0^*$ from $\mathcal{P}$
that best satisfies the latency and energy constraints,
and keep the accelerator, whose corresponding architecture $a_0^*$
has the highest accuracy, as the optimal accelerator.
Thus, we prove the effectiveness of our approach
if the architecture
$a_0^*\in\mathcal{P}$ corresponding
to the optimal accelerator found in Stage 2 produces (approximately) the same accuracy as $a^*$ obtained using the SOTA co-design.

In our experiment, we consider a target optimal accelerator $h^*$ as follows:
 512 PEs, NoC bandwidth constraint 900,
 off-chip bandwidth constraint 350, and KC-P dataflow.
In Fig.~\ref{fig: nasbench_pareto_3}, we plot all the optimal architectures under various latency and energy constraints.\footnote{MAESTRO returns the runtime cycles, instead of actual time,
for the inference latency.}
Then, we set three representative latency and energy consumption constraints, with their corresponding optimal models circled in red.
Next, we test each of the other 132 accelerators as the proxy, and find the corresponding set $\mathcal{P}$, which includes about $20$ optimal architectures
for that proxy. Then, we select the architecture from $\mathcal{P}$ whose
latency and energy are closest to the design constraints on the target
accelerator. We see that by using \emph{any} of the 132 accelerators
as the proxy, our  approach can still find the optimal architecture
that has (nearly) the same accuracy as that found by using SOTA \design.
Importantly, even  the proxy accelerator that has the lowest
SRCC with the target can yield an competitive architecture with
a good accuracy.

\subsubsection{Total search cost} We now compare
the total search cost incurred by exhaustive
search over our sampled space. Using the
coupled SOTA approach, the co-serach evaluates
133$*$1017$\approx$\textbf{135K} architecture-accelerator designs.
In Stage 1 of our approach, we choose one proxy and evaluate 1017 architectures to
obtain 20 optimal architectures for different latency and
energy constraints. As we use exhaustive search,
we do not need to run 20 times. In
Stage 2, we evaluate the remaining 132 accelerators combined
with the selected 20 architectures.
Thus, the total search cost of our approach is 132$*$20+1017$\approx$\textbf{3.7K}, which
is significantly less than 135K.
While reinforcement learning or evolutionary search
can improve the efficiency (especially on larger spaces), the order of the total cost remains
the same. Moreover,
when the architecture and accelerator spaces are larger,
the relative advantage of our approach is even more significant.

	\begin{table*}[!t]\normalsize
		\centering
		\begin{adjustbox}{width=0.9\textwidth}
			\begin{tabular}{ccccccccccc}
				\hline
				{\multirow{2}*{\textbf{\tabincell{c}{Accelerator \\ Index}}}}
				& \multicolumn{2}{c}{\textbf{SRCC}}
				
				&  \multicolumn{4}{c}{\textbf{Hardware Config.}} & \multicolumn{3}{c}{\textbf{Model Config.}}  \\
				& \textbf{Latency} & \textbf{Energy}  &  \textbf{PEs}  & \textbf{NoC}  & \textbf{Off-chip}  & \textbf{Dataflow}  & \textbf{Latency (cycles)}  & \textbf{Energy (nJ)} & \textbf{Accuracy (\%)}\\
				\hline
				\hline
				1 (\textbf{target}) & 1 & 1 & 512 &  900 & 350 & KC-P & 2061611 & 614779 & 69.60 \\
				\cdashline{1-11}[0.8pt/2pt]
				64 & 0.638 & 0.945 & 512 &  400 & 350 & X-P &   2061611 & 602782 & 69.58 \\
				\cdashline{1-11}[0.8pt/2pt]
				91  & 0.775 & 0.945 & 32 &  800 & 250 & X-P & 2046476 & 610891 & 69.60 \\
				\hline
				1 (\textbf{target}) & 1 & 1 & 512 &  900 & 350 & KC-P &   3367489 & 965462 & 71.18\\
				\cdashline{1-11}[0.8pt/2pt]
				64 & 0.638 & 0.945  & 512 &  400 & 350 & X-P &   3367489 & 965462 & 71.18\\
				\cdashline{1-11}[0.8pt/2pt]
				91  & 0.775 & 0.945  & 32 &  800 & 250 & X-P &   3367489 & 965462 & 71.18\\
				\hline
				1 (\textbf{target}) & 1 & 1 & 512 &  900 & 350 & KC-P &   5923046 & 1858261 & 71.76\\
				\cdashline{1-11}[0.8pt/2pt]
				64 & 0.638 & 0.945  & 512 &  400 & 350 & X-P &   5923046 & 1858261 & 71.76\\
				\cdashline{1-11}[0.8pt/2pt]
				91  & 0.775 & 0.945  & 32 &  800 & 250 & X-P &   5923046 & 1858261 & 71.76\\
				\hline
			\end{tabular}
		\end{adjustbox}
		\vspace{0.2cm}
		\caption{\note{Hardware configuration of the target and two proxy accelerators, and performance metrics of the selected optimal models on each of them.
				``Accelerator Index" corresponds to the $x$-axis in right of Fig.~\ref{fig: alphanet_pareto_3}, models on the target accelerator correspond to the circled ones in left of Fig.~\ref{fig: alphanet_pareto_3}, while the selected optimal models on proxy accelerators correspond to the diamond marks locating on the accelerator indexes. The architecture configuration of the target models is further illustrated in Table~\ref{table: alphanet_pareto_3_config2}.}}
		\label{table: alphanet_pareto_3_config1}
	\end{table*}

\begin{table*}[!t]\normalsize
	\centering
	\begin{adjustbox}{width=0.9\textwidth}
		\begin{tabular}{cccccc}
			\hline
			{\multirow{2}*{\textbf{\tabincell{c}{Target Model}}}}  &
			\multicolumn{5}{c}{\textbf{Model Architecture}}  \\
			& \textbf{Resolution} & \textbf{Width}  &  \textbf{Kernel Size}  & \textbf{Expansion Ratio} & \textbf{Depth}\\
			\hline
			\hline
			\#1 & 224 & 16, 16, 24, 32, 64, 112, 192, 216, 1792 &  3, 3, 3, 3, 3, 3, 3 & 1, 4, 4, 6, 6, 5, 6 & 1, 3, 4, 3, 3, 3, 1 \\
			\cdashline{1-6}[0.8pt/2pt]
			
			\#2 & 288 &  16, 16, 24, 32, 64, 112, 192, 216, 1792 &  3, 3, 3, 3, 3, 7, 3 & 1, 4, 4, 5, 4, 5, 6 & 1, 3, 3, 3, 4, 4, 1 \\
			\cdashline{1-6}[0.8pt/2pt]
			
			\#3 & 288 & 16, 16, 24, 32, 64, 112, 192, 216, 1792 &  3, 3, 5, 7, 7, 7, 3 & 1, 6, 6, 6, 5, 5, 6 & 1, 6, 6, 3, 6, 6, 1 \\
			\cdashline{1-6}[0.8pt/2pt]
			
			
			\hline
		\end{tabular}
	\end{adjustbox}
	\vspace{0.2cm}
	\caption{\note{Architecture configuration of target models in Fig.~\ref{fig: alphanet_pareto_3}. The first row (i.e., target model \#1) corresponds to the leftmost circled model in Fig.~\ref{fig: alphanet_pareto_3}, and second row corresponds to the middle circled model, etc. For detailed explanation of the operations in AlphaNet search space, please refer to \cite{alphanet_attentive_nas}.}}
	\label{table: alphanet_pareto_3_config2}
\end{table*}

\begin{figure}[!t]\centering
	
	\subfigure[SRCC of Latency]{
		\centering
		\includegraphics[width=0.295\linewidth]{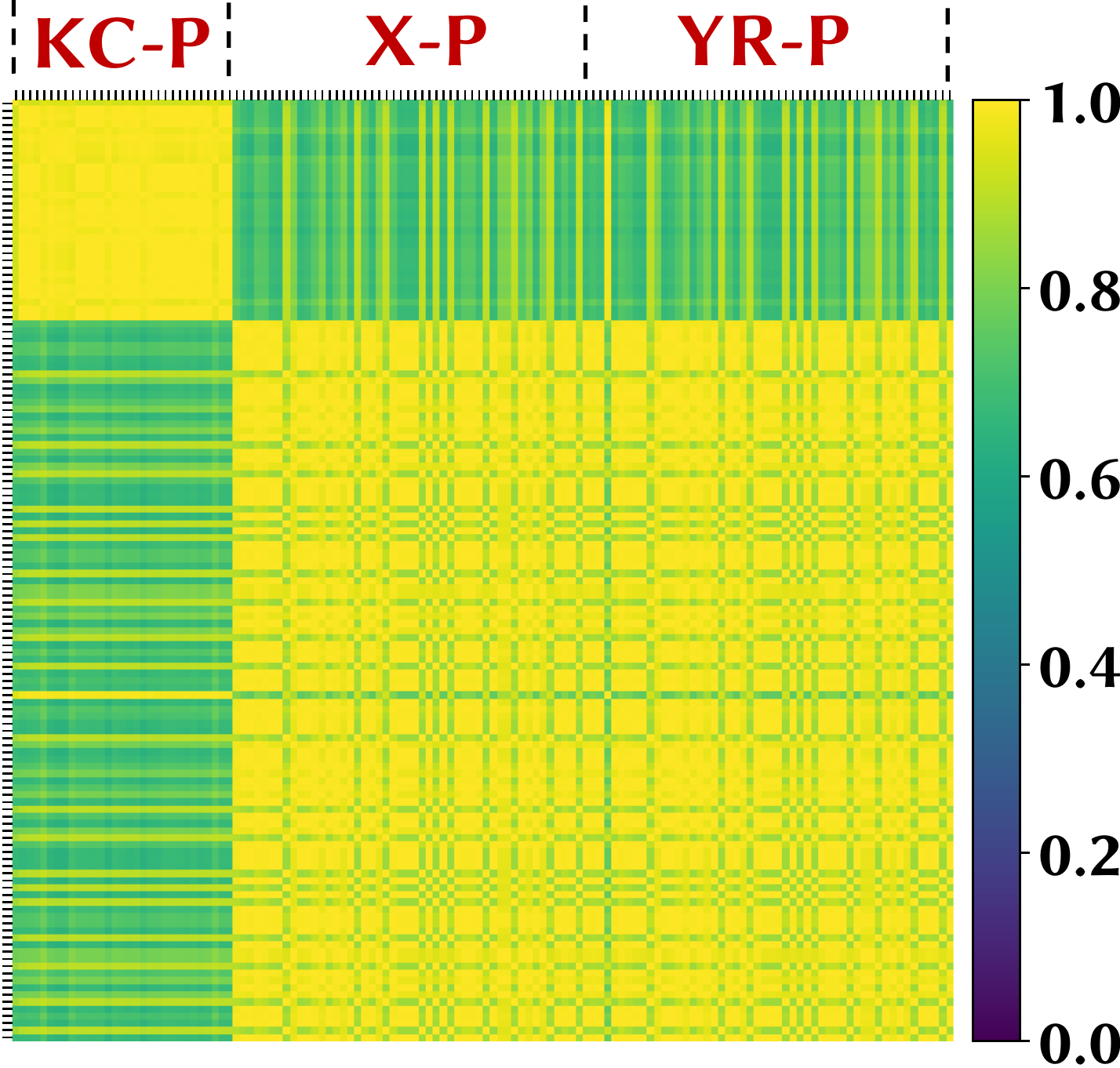}\vspace{-0.6cm}
		\label{fig: alphanet_latency}
	}
	\subfigure[SRCC of Energy Consumption]{
		\centering
		\includegraphics[width=0.295\linewidth]{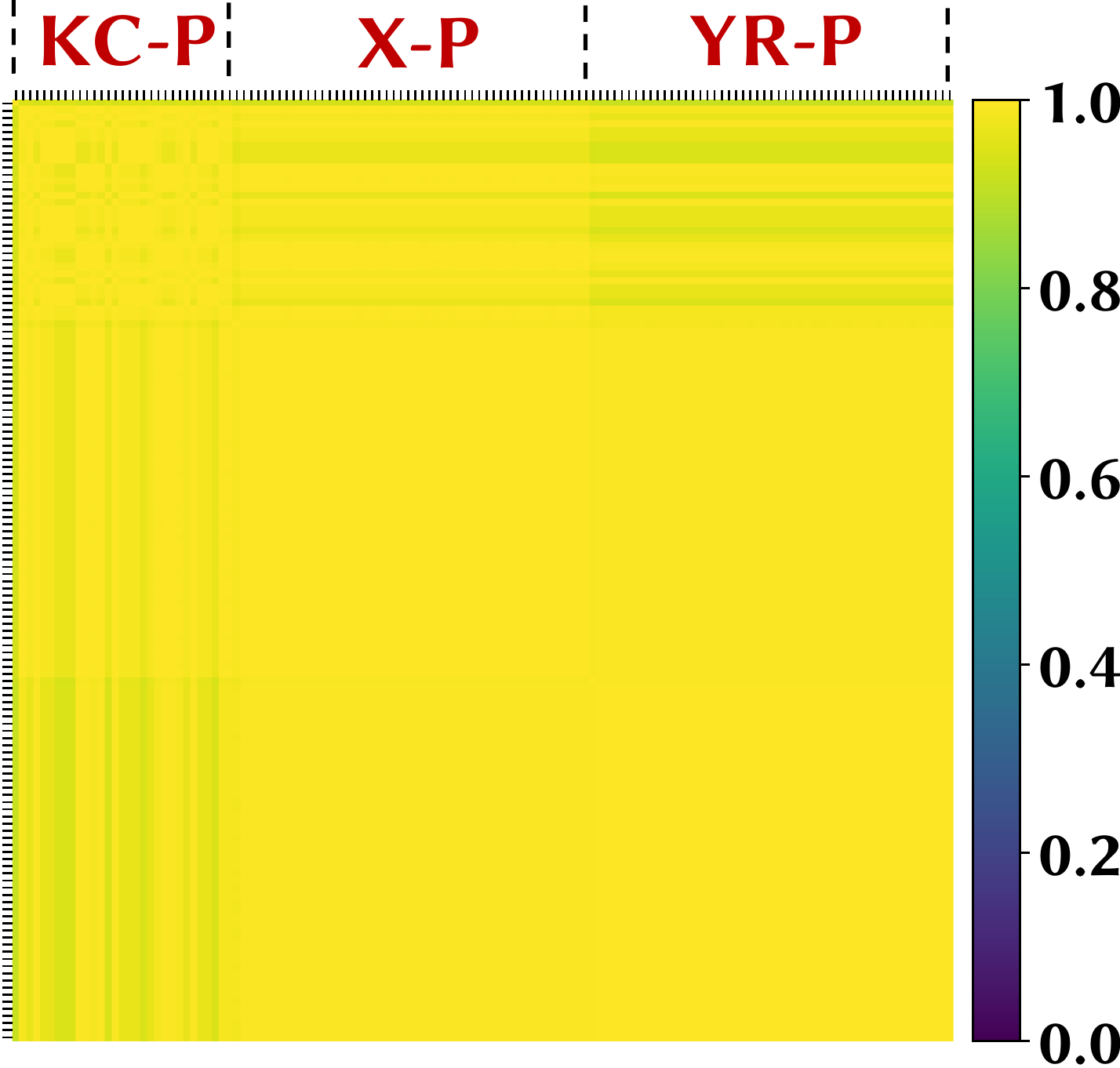}\vspace{-0.6cm}
		\label{fig: alphanet_energy}
	}
	\subfigure[CDF of SRCC]{
		\centering
		\includegraphics[width=0.33\linewidth]{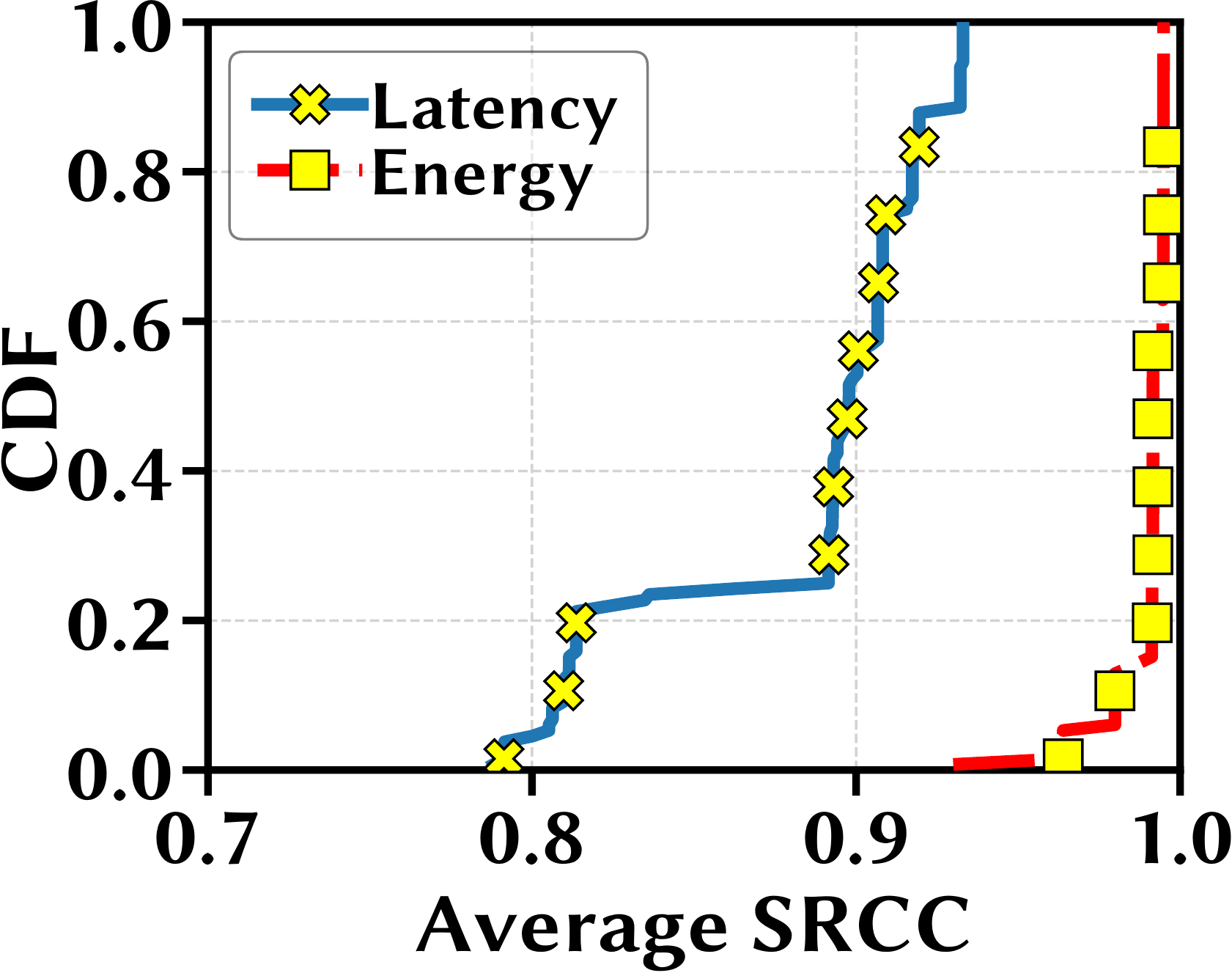}\vspace{-0.6cm}
		\label{fig: alphanet_cdf}
	}
	\vspace{-0.4cm}
	\caption{Performance monotonicity. We test 1046 models sampled in AlphaNet search space on 132 accelerators.}
	\vspace{-0.3cm}
	\label{fig: alphanet_srcc}
\end{figure}

\subsection{AlphaNet}

\begin{figure}[!t]
	\centering
	\includegraphics[width=0.48\textwidth]{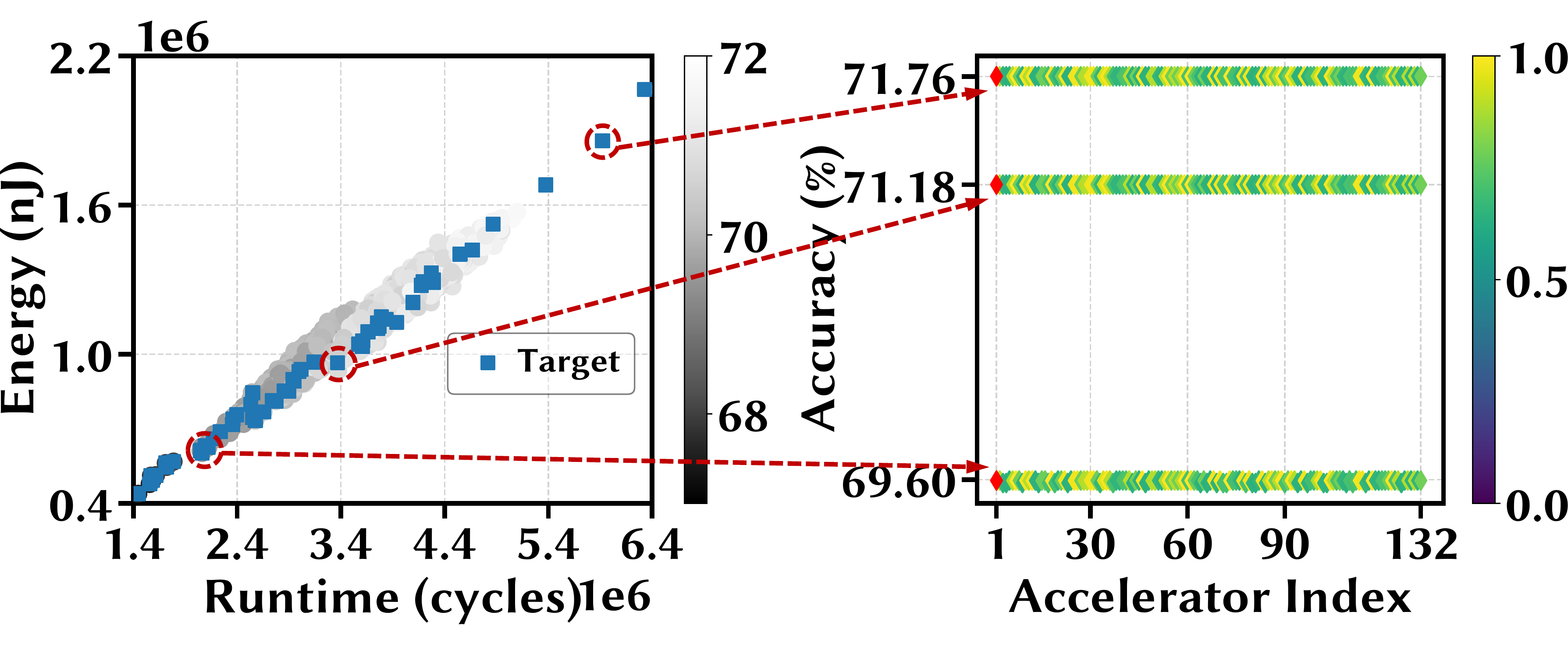}
\vspace{-0.6cm}	
\caption{AlphaNet. Left: The optimal
models are marked in blue, and the grey scale indicates accuracy.
Right: The accuracy of the model selected
from the proxy's optimal model set. We test each accelerator
as a different proxy.
\note{We select two proxy accelerators (indexes 64 and 91) and show the detailed results in Table~\ref{table: alphanet_pareto_3_config1}.}
}
	\label{fig: alphanet_pareto_3}
\vspace{-0.3cm}
\end{figure}

We now turn to the AlphaNet architecture space,
and show the results in Fig.~\ref{fig: alphanet_srcc}
and Fig.~\ref{fig: alphanet_pareto_3}.
While the SRCC values are lower than those
in the NAS-Bench-301 case, they are still generally very high
(e.g., mostly >0.9). Crucially, as shown in Fig.~\ref{fig: alphanet_pareto_3},
our  approach can successfully find an architecture that has
(almost) the same accuracy as that obtained by using the SOTA coupled approach.

\begin{figure}[!t]\centering
	
	\subfigure[Latency SRCC]{
		\centering
		\includegraphics[width=0.295\linewidth]{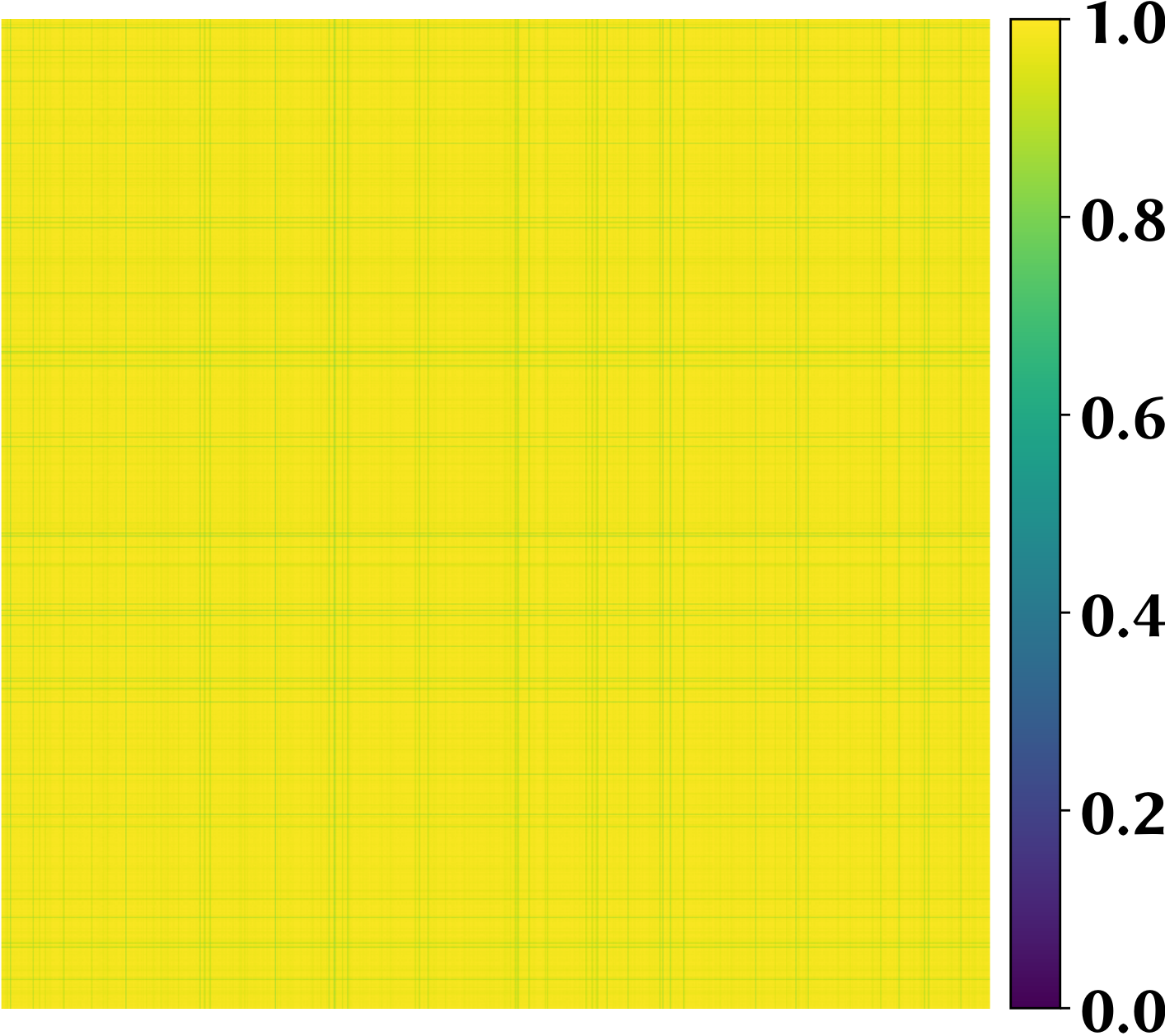}\vspace{-0.6cm}
		\label{fig: nasbench_latency_mix}
	}
	\subfigure[Energy SRCC]{
		\centering
		\includegraphics[width=0.295\linewidth]{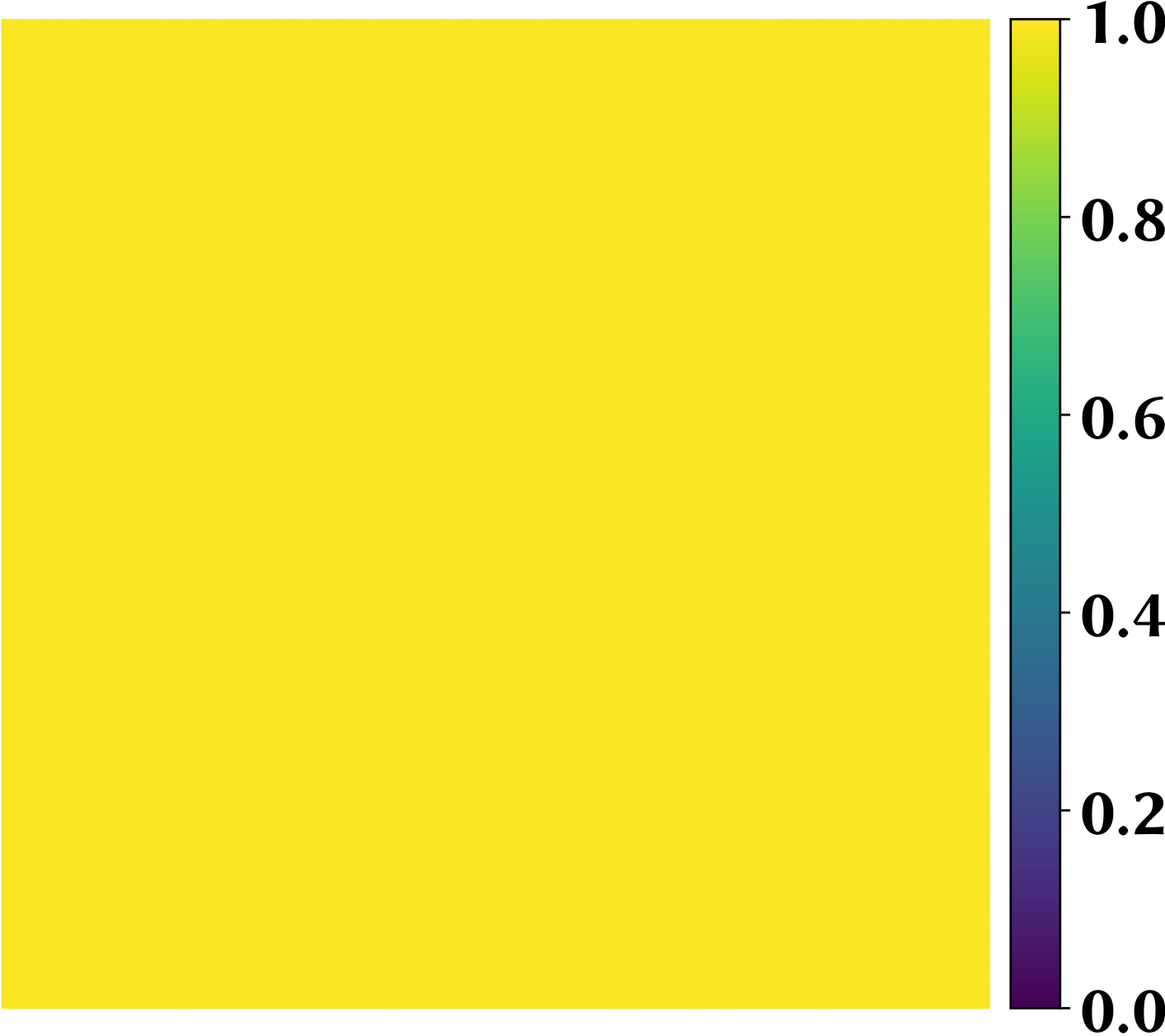}\vspace{-0.6cm}
		\label{fig: nasbench_energy_mix}
	}
	\subfigure[CDF of SRCC]{
		\centering
		\includegraphics[width=0.33\linewidth]{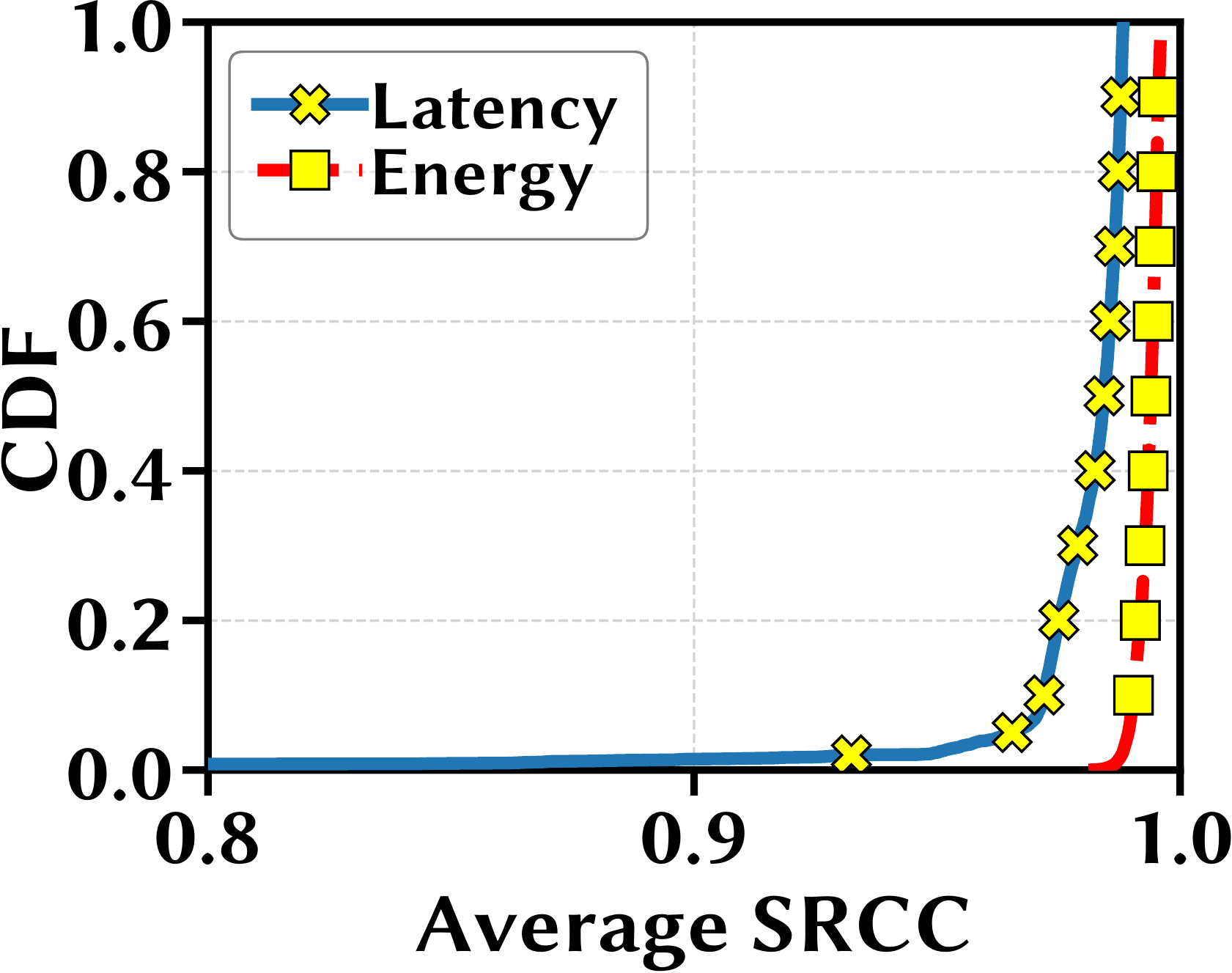}\vspace{-0.6cm}
		\label{fig: nasbench_cdf_mix}
	}
	\vspace{-0.4cm}
	\caption{Performance monotonicity. We test 1017 models sampled in DARTS on 5000 accelerators with layer-wise mixed dataflows.}
	\vspace{-0.3cm}
	\label{fig: nasbench_mix}
\end{figure}

\begin{figure}[!t]\centering
	
	\subfigure[Latency SRCC]{
		\centering
		\includegraphics[width=0.295\linewidth]{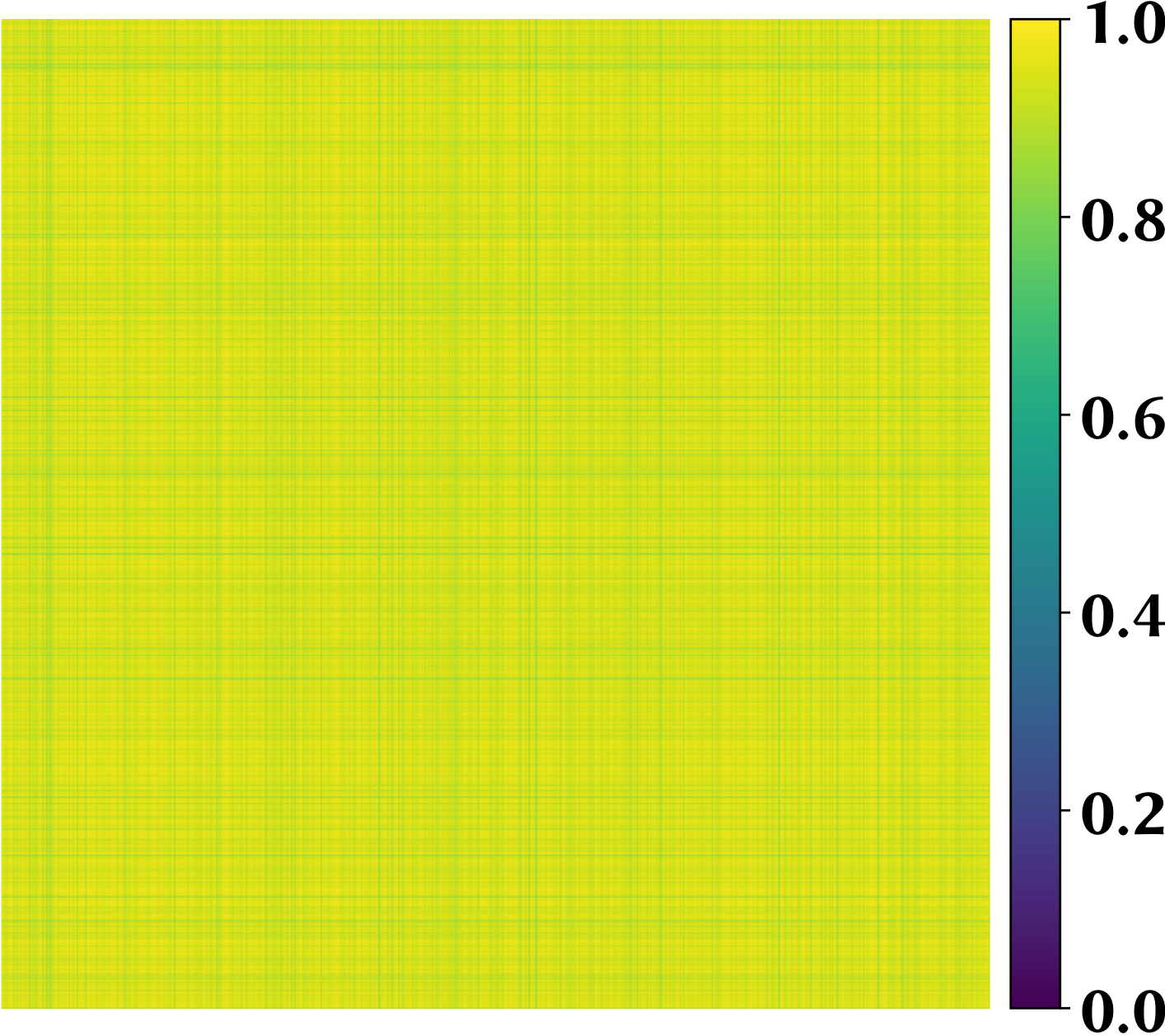}\vspace{-0.6cm}
		\label{fig: alphanet_latency_mix}
	}
	\subfigure[Energy SRCC]{
		\centering
		\includegraphics[width=0.295\linewidth]{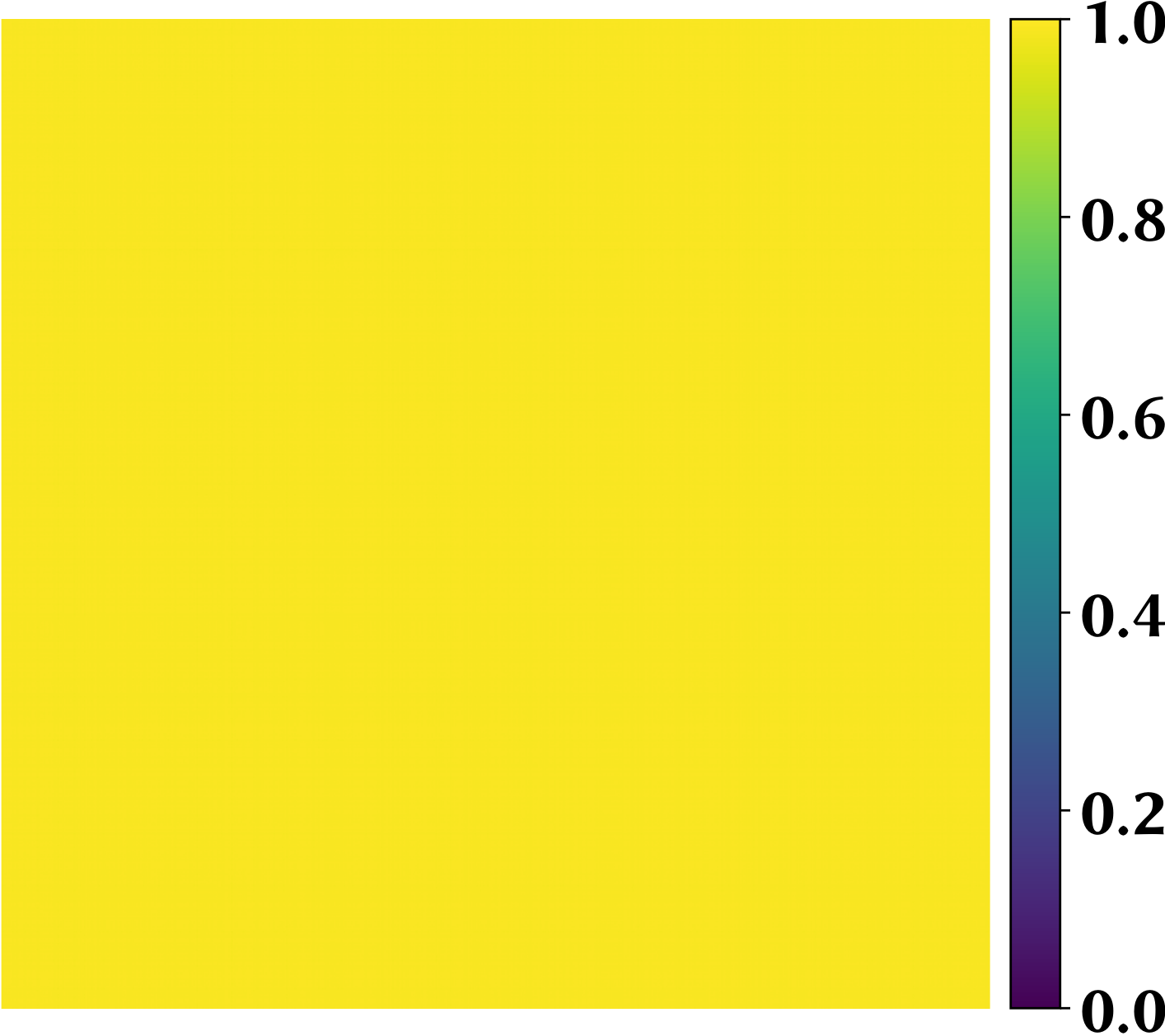}\vspace{-0.6cm}
		\label{fig: alphanet_energy_mix}
	}
	\subfigure[CDF of SRCC]{
		\centering
		\includegraphics[width=0.33\linewidth]{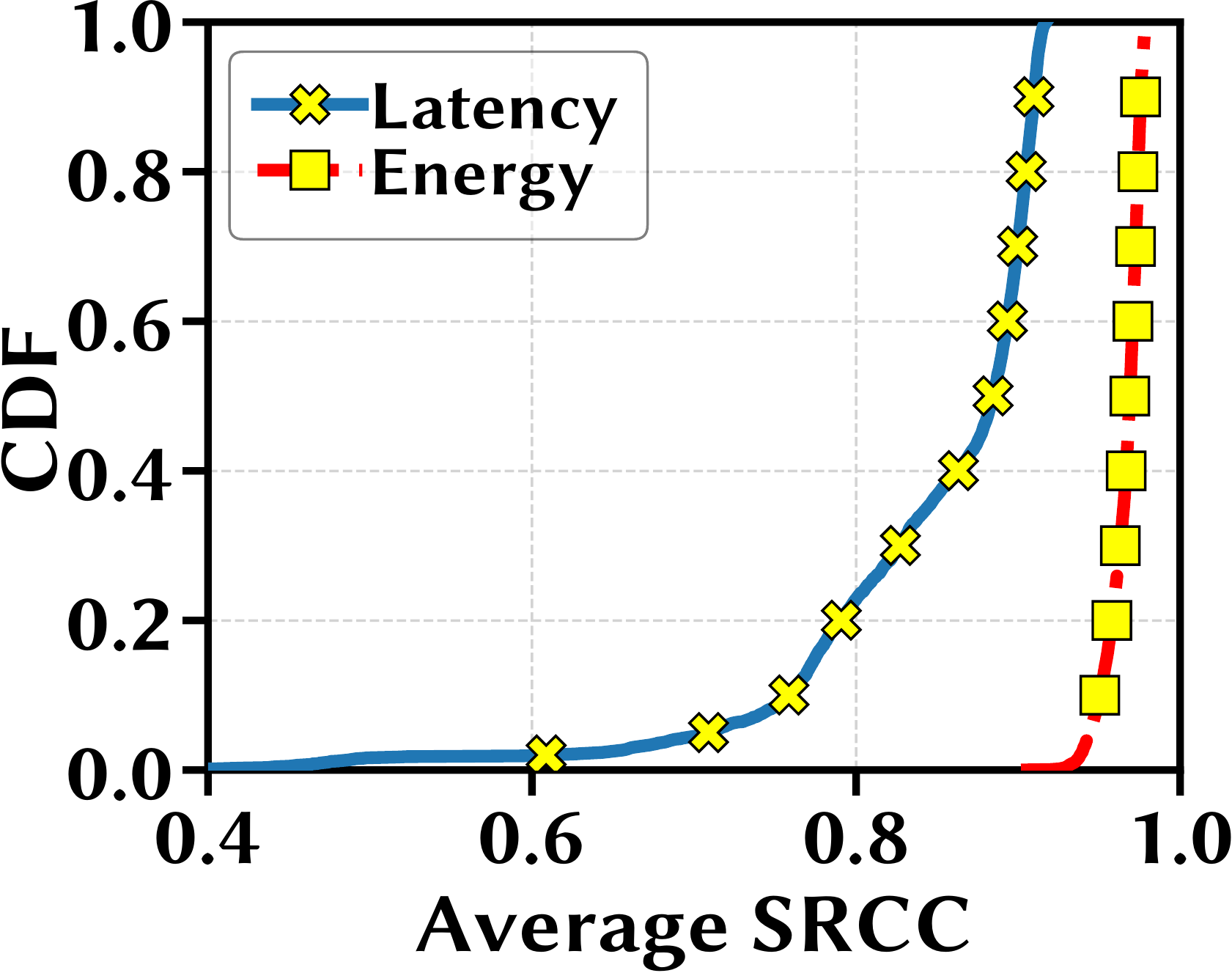}\vspace{-0.6cm}
		\label{fig: alphanet_cdf_mix}
	}
	\vspace{-0.4cm}
	\caption{Performance monotonicity. We test 1046 models sampled in AlphaNet on 5000 accelerators with layer-wise mixed dataflows.}
	\vspace{-0.3cm}
	\label{fig: alphanet_mix}
\end{figure}

\subsection{Layer-wise Mixed Dataflow}\label{sec:experiment_mixing}
Ideally,
each layer of a DNN model can be switched between accelerator hardware and dataflows to search for the best combination (especially in
the multi-accelerator design case) \cite{yiyu_codesign_dac}.
To account for this, we divide each model into 22 parts: first and last convolutional layer, and evenly into 20 groups for all intermediate layers.
For each part, it can be executed on any of our 51 sampled hardware configurations
following any dataflow.
We sample 5000 different mixtures for our models in NAS-Bench-301 and AlphaNet spaces,
 and report the SRCC results in Fig.~\ref{fig: nasbench_mix}
and~\ref{fig: alphanet_mix}, respectively. The results
confirm again that strong performance monotonicity exists
and ensures the effectiveness of our  approach.
We omit the optimal accuracy results due to the lack
of space, while noting that they are similar to Figs.~\ref{fig: nasbench_pareto_3}
and~\ref{fig: alphanet_pareto_3}.

\section{Related Work}\label{sec: related_work}

\textbf{NAS and accelerator design.}
Hardware-aware NAS has been actively studied to incorporate characteristics of target device
and automate the design of optimal architectures
subject to latency and/or energy constraints \cite{DNN_NAS_Hardware_YiyuShi_ICCAD_2019_Jiang:2019:AVE:3316781.3317757,DNN_NAS_HardwareNAS201_Benchmark_Rice_ICLR_2021_li2021hwnasbench,DNN_NAS_NetAdapt_ECCV_MIT_2018_10.1007/978-3-030-01249-6_18,weight_sharing_perform_random_search_CVPR, multi_hardware_mobile_models_NAS_CVPR, NAS_survey_arxiv,DNN_NAS_MnasNet_Google_CVPR_2019_Tan_2019, DNN_FBNet_HardwareAwareConvNetDesign_CVPR_2019_Wu2018FBNetHE}.
These studies do not explore the hardware design space.
A recent NAS study \cite{Shaolei_NAS_OneProxy_Sigmetrics_Journal_Dec2021}
explores latency monotonicity to scale up NAS across different devices,
but it only considers latency constraints and, like
other NAS studies, does explore the hardware design space.
In parallel, there have also been studies on automating the design
of accelerators for DNNs \cite{DNN_AutoDNNChip_FPGA_ASIC_YingyanLin_Rice_DemingChen_UIUC_FPGA_2020_10.1145/3373087.3375306}.
But, NAS and accelerator design have been traditionally studied in a siloed manner,
resulting in sub-optimal designs.

\textbf{Architecture-accelerator co-design.}
The studies on jointly optimizing architectures and accelerators have
been quickly expanding.
For example,
\cite{yiyu_codesign_dac} jointly optimizes neural architectures and ASIC accelerators
using reinforcement learning, \cite{yiyu_co_exploration} performs a two-level (fast and slow) hardware exploration for each candidate neural architecture,
\cite{DNN_NAS_StandingShoulder_YiyuShi_HardwareCoDesign_TCAD_2020_jiang2020standing} adopts a set of manually selected models as the hot start state for acceleration exploration,
and \cite{song_han_neural_accelerator_architecture_search}
 co-designs
neural architecture, hardware configuration and dataflow,
and employs evolutionary search to reduce the search cost.
These studies primarily focus on improving the
search efficiency given a certain search space.
By contrast,
we use a principled approach
to reducing the total search space, without losing optimality.

\pdfoutput=1

\section{Conclusion}
\label{sec: conclusion}

In this paper, we reduce the total
\design cost by semi-decoupling NAS from accelerator design.
Concretely, we demonstrate latency and energy  monotonicity among different accelerators, and use just one proxy accelerator's optimal architecture set to avoid searching
over the entire architecture space.
Compared to the SOTA co-designs, our approach can reduce the total design complexity by orders of magnitude, without losing optimality. Finally, we validate our approach via experiments
on two
search spaces --- NAS-Bench-301 and AlphaNet.

\section*{Acknowledgement}
B. Lu and S. Ren were supported in part by the U.S. National Science
Foundation under grant CNS-1910208. Z. Yan and S. Shi were supported
in part by the U.S. National Science Foundation under grant CNS-1822099.

\bibliographystyle{ACM-Reference-Format}

\end{document}